\documentstyle[12pt,epsfig]{article}
\renewcommand{\bar}[1]{\overline{#1}}

\begin{document}
\bigskip\bigskip
\begin{flushright}
GEF-Th-6/2010\\
April 2010
\end{flushright}

\begin{center}
{\large \bf Deep Inelastic Processes\\
and the Equations of Motion}
\end{center}

\vspace{12pt}

\begin{center}
 {\bf E.~Di~Salvo\footnote{Elvio.Disalvo@ge.infn.it}\\}

 {Dipartimento di Fisica and I.N.F.N. - Sez. Genova, Via Dodecaneso, 33 \\-
 16146 Genova, Italy\\}

\end{center}

\vspace{10pt}
\begin{center} {\large \bf Abstract}

We show that the Politzer theorem on the equations of motion implies 
approximate constraints on the quark correlator. These, in turn, restrict considerably, for sufficiently large $Q^2$, the number of independent distribution functions that characterize the internal structure of the nucleon, and of independent fragmentation functions. This result leads us to suggesting an alternative method for determining transversity. Moreover our approach
implies predictions on the $Q^2$-dependence of some azimuthal asymmetries, like Sivers, Qiu-Sterman and Collins asymmetry. Lastly, we discuss some implications on the Burkhardt-Cottingham sum rule.
\end{center}

\vspace{10pt}

\centerline{PACS Numbers: 13.85.Ni, 13.88.+e, 11.15.-q}

\newpage


\section{Introduction}
The problem of calculating inclusive cross sections at high energies and high momentum transfers has become quite important in the last two decades, during which a lot of experimental data on deep inelastic processes have been 
accumulated. In particular we refer to deep inelastic scattering (DIS) (Ashman {\it et al.}, 1988, 1989; Adeva {\it et al.}, 1998; Anthony {\it et al.}, 1996a,b, 2003; Abe {\it et al.}, 1997a,b, 1998; Airapetian {\it et al.}, 1998; Yun {\it et al.}, 2003; Zheng {\it et al.}, 2004), semi-inclusive DIS (SIDIS) (Arneodo {\it et al.}, 1987; Ashman {\it et al.}, 1991; Adams {\it et al.}, 1993; Airapetian {\it et al.}, 2000, 2001,2003, 2005a,b; Diefenthaler, 2005; Bravar {\it et al.}, 1999; Alexakhin {\it et al.}, 2005; Ageev {\it et al.}, 2007; Bressan, 2007; Avakian {\it et al.}, 2005; Alekseev {\it et al.}, 2010a,b), Drell-Yan (DY) (Falciano {\it et al.}, 1986; Guanziroli {\it et al.}, 1988; Conway {\it et al.}, 1989; McGaughey {\it et al.}, 1994; Towell {\it et al.}, 2001; Zhu {\it et al.}, 2007), $e^+e^-$ annihilation into two back-to-back jets (Abe {\it et al.}, 2006), while analogous experiments have been planned recently (Bunce {\it et al.}, 2000; Lenisa and Rathmann, PAX Coll., Julich, hep-ex/0505054, 2005; Lenisa, 2005; Afanasev {\it et al.}, Jefferson Lab., hep-ph/0703288, 2007; Hawranek, 2007; Kotulla {\it et al.}, Technical Progress Report for PANDA, 2005). One of the aims of high energy physicists is to extract from data distribution and/or fragmentation functions, especially if unknown. Among them, the transversity (Ralston and Soper, 1979; Artru and Mekhfi, 1990; Jaffe and Ji, 1991a, 1992) is of particular interest, since it is the only twist-2 distribution function for which very poor information (Soffer, 1995; Anselmino {\it et al.}, 2007) is available till now. But also transverse momentum dependent (TMD) functions - especially the T-odd ones - are taken in great consideration; for instance, knowledge of the Collins (1993) fragmentation function or of the Boer-Mulders (1998) function could help to extract transversity, which is chiral-odd and therefore couples only with chiral-odd functions. Moreover, TMD functions are involved in several intriguing azimuthal asymmetries, like the already mentioned Collins (1993) and Boer-Mulders (1998) effects, or the Sivers (1990, 1991), Qiu-Sterman (1991, 1992, 1998) and Cahn (1978, 1989) effects, which, in part, have found experimental confirmation (Airapetian {\it et al.}, 2005a,b; Diefenthaler, 2005; Bravar {\it et al.}, 1999; Alexakhin {\it et al.}, 2005; Bressan, 2007; Abe {\it et al.}, 2006)  and, in any case, have stimulated a great deal of articles (Mulders and Tangerman, 1996; Boer {\it et al.}, 2000, 2003a,b; Brodsky {\it et al.}, 2002a,b, 2003; Di Salvo, 2007a; Collins {\it et al.}, 2006; Efremov {\it et al.}, 2006a,b, 2009; Avakian {\it et al.}, 2008a,b; Boffi {\it et al.}, 2009; Anselmino {\it et al.}, 2009a,b, 2010; Boer, 2009). Last, some questions remain open, among which the parton interpretation of the polarized structure function $g_2$ (Anselmino {\it et al.}, 1995; Jaffe and Ji, 1991a). Obviously, all of these data and kinds of problems are confronted with the QCD theory and in this comparison short and long distance scales are interested, so that the factorization theorems (Collins, 1998, 1989; Collins {\it et al.}, 1988; Sterman, 2005) play quite an important role in separating the two kinds of effects. Strong contributions in this sense have been given by Politzer (1980), Ellis {\it et al.} (1982, 1983) (EFP), Efremov and Radyushkin (1981), Efremov and Teryaev (1984), Collins and Soper (1981, 1982), Collins {\it et al.} (1988) and Levelt and Mulders (1994) (LM).

In the present paper we propose an approach somewhat similar to EFP's and to LM's, but we use more extensively the Politzer (1980) theorem on equations of motion (EOM). We consider in particular the hadronic tensor for SIDIS, DY and $e^+e^- \to \pi \pi X$. We also consider energies and momentum transfers high enough for assuming one photon approximation, but not so large that weak interactions be comparable with electromagnetic ones. As regards time-like 
photons, we assume to be far from masses of vector resonances, like $J/\Psi$, $\Upsilon$ or $Z^0$. Lastly, we do not consider the case of active (anti-)quarks originating from gluon annihilation.   

Our starting point is the "Born" (LM) approximation for the hadronic tensor, which reads, in the three above mentioned reactions, as
\begin{equation}
W_{\alpha\beta}(P_A, P_B, q) = C \sum_ae^2_a\int \frac{d^4p}{(2\pi)^4} Tr\left[\Phi^a_A(p) 
\gamma_{\alpha}\Phi^b_B(p') \gamma_{\beta}\right]. \label{ht00} 
\end{equation}  
Here $C$ is due to color degree of freedom, $C = 1$ for SIDIS and $1/3$ for DY and $e^+e^-$ annihilation. $p$ and $p'$ 
denote the four-momenta of the active partons, such that 
\begin{equation}
p \mp p' = q, \label{mom} 
\end{equation} 
$q$ being the four-momentum of the virtual photon and the $-$ sign referring to SIDIS, the $+$ to DY or to $e^+e^-$ annihilation. $\Phi_A$ and $\Phi_B$ are correlators, relating the active partons to the (initial or final) hadrons $h_A$ and $h_B$, whose four-momenta are, respectively, $P_A$ and $P_B$. We restrict 
ourselves to spinless and spin-1/2 hadrons. $a$ and $b$ are the flavors of the active partons, with $a = u, d, s, \bar{u}, \bar{d}, \bar{s}$ and $b = a$ in SIDIS, $b = \bar{a}$ in DY and $e^+e^-$ annihilation; $e_a$ is the fractional 
charge of flavor $a$. In DY $\Phi_A$ and $\Phi_B$ encode information on the active quark and antiquark distributions inside the initial hadrons. In SIDIS $\Phi_B$ is replaced by the fragmentation correlator $\Delta_B$, describing the 
fragmentation of the struck quark into the final hadron $h_B$. In the case of $e^+e^-$ annihilation, both correlators $\Phi_A$ and $\Phi_B$ have to be replaced by $\Delta_A$ and $\Delta_B$ respectively. 

In the approximation considered we define the distribution correlator (commonly named correlator) as
\begin{equation}
\Phi_{ij}(p; P,S) = N\int\frac{d^4x}{(2\pi)^4} e^{ipx} 
\langle P,S|\bar{\psi}_j(0) \psi_i(x)|P,S\rangle. \label{corr0}
\end{equation}
Here $N$ is a normalization constant, to be determined in sect. 4. $\psi$ is the 
quark\footnote{For an antiquark eqs. (\ref{corr0}) and (\ref{corrp}) should be slightly modified, as we shall see in 
sects. 2 and 6.} field of a given flavor and $|P,S\rangle$ a state of a hadron (of spin 0 or 1/2) with a given 
four-momentum $P$ and Pauli-Lubanski (PL) four-vector $S$, while $p$ is the quark four-momentum. The color and flavor 
indices have been omitted in $\psi$ for the sake of simplicity and from now on will be forgotten, unless differently 
stated. On the other hand, the fragmentation correlator is defined as
\begin{equation}
\Delta_{ij}(p; P,S) = N\int\frac{d^4x}{(2\pi)^4} e^{ipx} 
\langle 0|\bar{\psi}_j(0)a(P,S)a^\dagger(P,S)\psi_i(x)|0\rangle, \label{corrp}
\end{equation}
where $a(P,S)[a^\dagger(P,S)]$ is the destruction (creation) operator for the fragmented hadron, of given four-momentum and PL four-vector. 

The hadronic tensor (\ref{ht00}) is not color gauge invariant. Introducing a gauge link is not sufficient to fulfil this condition, but EOM suggest to add suitable contributions of higher correlators, involving two quarks and a number 
of gluons, so as to construct a gauge invariant hadronic tensor. 

We adopt an axial gauge, obtaining for the correlator a $gM/Q$ expansion, where $g$ is the coupling, $M$ the rest mass of the hadron and $Q$ the QCD "hard" energy scale, generally assumed equal to $\sqrt{|q^2|}$. We examine in detail the first two terms of the expansion. The zero order term corresponds to the QCD parton model approximation. As regards the second term, it concerns the T-odd functions; in particular, we discuss an interesting approximation, already proposed by Collins (2002). In both cases we obtain several approximate relations among "soft" functions, which survive perturbative QCD evolution, as a consequence of EOM. Our approach allows also to determine the $Q$-dependence of some important azimuthal asymmetries and to draw conclusions about the Burkhardt-Cottingham (1970) sum rule. 
 
 Section 2 is devoted to the gauge invariant correlator (more properly to the distribution correlator), whose properties are deduced with the help of EOM. In particular, we derive an expansion in powers of $gM/Q$, whose terms can be interpreted as Feynman-Cutkosky graphs. In section 3 we give a prescription for writing a gauge invariant sector of the hadronic tensor which is of interest for interactions at high $Q$. In sects. 4 and 5 we study in detail the zero 
order term and the first order correction of the expansion, deducing approximate relations among functions which appear in the usual parameterizations of the correlator (Mulders and Tangerman, 1996; Goeke {\it et al.}, 2005). Sect. 6 is dedicated to the fragmentation correlator. In sect. 7 we illustrate the azimuthal asymmetries involved in the three different deep inelastic processes. Lastly sect. 8 is reserved to a summary of the main results of the paper. 

\section{Gauge Invariant Correlator}

The correlator (\ref{corr0}) can be made gauge invariant, by inserting between the quark fields a link operator (Collins and Soper, 1981, 1982; Mulders and Tangerman, 1996), in the following way:
\begin{equation}
\Phi_{ij}(p; P,S) = N\int\frac{d^4x}{(2\pi)^4} e^{ipx} 
\langle P,S|\bar{\psi}_j(0) {\cal L}(x)  \psi_i(x)|P,S\rangle. \label{corr1}
\end{equation}
 Here
\begin{equation}
{\cal L}(x) = {\mathrm P} exp\left[ig\Lambda_{\cal I}(x)\right], \ ~~~~~ \ {\mathrm with}
\ ~~~~~ \ \Lambda_{\cal I}(x) = \int_{0({\cal I})}^x \lambda_a A^a_{\mu}(z)dz^{\mu}, \label{link}
\end{equation}
is the gauge link operator, "P" denotes the path-ordered product along a given integration contour ${\cal I}$, 
$\lambda_a$ and $A^a_{\mu}$ being respectively the Gell-Mann matrices and the gluon fields. The link operator depends on the choice of ${\cal I}$, which has to be fixed so as to make a physical sense. According to previous 
treatments (Mulders and Tangerman, 1996; Collins, 2002; Boer {\it et al.}, 2003b; Bomhof {\it et al.}, 2004), we define two different contours, ${\cal I}_{\pm}$, as sets of three pieces of straight lines, from the origin to $x_{1\infty}\equiv (\pm\infty, 0, {\bf 0}_{\perp})$, from $x_{1\infty}$ to 
$x_{2\infty}\equiv (\pm\infty, x^+, {\bf x}_{\perp})$ and from $x_{2\infty}$ to  $x\equiv (x^-, x^+,{\bf x}_{\perp})$, having adopted a frame, whose $z$-axis is taken along the hadron momentum, with $x^{\pm} = 1/\sqrt{2}(t\pm z)$. We remark that the choice of the path is important for the so-called T-odd\footnote{More precisely, one should speak of "naive T", consisting of reversing all momenta and angular momenta involved in the process, without interchanging initial and final states(DeRujula, 1971; Bilal {\it et al.}, 1991; Sivers, 2006).} functions (Boer and Mulders, 1998): the path ${\cal I}_+$ is suitable for DIS distribution functions, while ${\cal I}_-$ has to be employed in DY (Boer {\it et al.}, 2003b; Bomhof {\it et al.}, 2004). For an antiquark the signs of the correlator (\ref{corr1}) and of the four-momentum $p$ have to be changed.
 
In the following of the section we investigate some properties of the correlator. 
  
\subsection{T-even and T-odd correlator}
  
We set (Boer {\it et al.}, 2003b)
\begin{equation}
\Phi_{E(O)} = \frac{1}{2}[\Phi_+\pm\Phi_-], \label{spl}
\end{equation}
where $\Phi_{\pm}$ corresponds to the contour ${\cal I}_{\pm}$ in eqs. (\ref{link}), while $\Phi_E$ and $\Phi_O$ select respectively the T-even and the T-odd "soft" functions. These two correlators contain respectively the link operators ${\cal L}_E(x)$ and ${\cal L}_O(x)$, where
\begin{equation}
{\cal L}_{E(O)}(x) = \frac{1}{2} {\mathrm P} \left\{exp\left[ig\Lambda_{{\cal I}_+}(x)\right]\pm 
exp\left[ig\Lambda_{{\cal I}_-}(x)\right]\right\} \label{spl1}
\end{equation}
and $\Lambda_{{\cal I}_{\pm}}(x)$ are defined by the second eq. (\ref{link}). Eqs. (\ref{spl}) and (\ref{spl1}) imply that the T-even functions are independent of the contour (${\cal I}_+$ or ${\cal I}_-$), while the T-odd ones change sign according as to whether they are involved in DIS or in DY (Collins, 2002; Boer {\it et al.}, 2003b). In this sense, such functions are not strictly 
universal (Collins, 2002), as already stressed. It is convenient to consider an axial gauge, 
\begin{equation}
{\bf A}^+ = 0, \label{axi}
\end{equation}
with antisymmetric boundary conditions (Mulders and Tangerman, 1996): 
\begin{equation}
{\bf A}^{\mu}(-\infty, x^+, {\bf x}_{\perp}) = -{\bf A}^{\mu}(+\infty, x^+, {\bf x}_{\perp}). 
\end{equation} 

Here we have adopted the shorthand notation ${\bf A}^{\mu}$ for $\lambda^a A_a^{\mu}$. In this gauge - proposed for the 
first time by Kugut and Soper (1970) and named KS gauge in the following - we have
\begin{equation}
\Lambda_{{\cal I}_+}(x) = -\Lambda_{{\cal I}_-}(x) = 
\int_{x_1}^{x_2} dz_{\mu}{\bf A}^{\mu}(z)\label{link0},
\end{equation}
where $x_i$ is a shorthand notation for $x_{i,+\infty}$, $i = 1, 2$. Therefore, in the KS gauge,
\begin{equation}
{\cal L}_E(x) = {\mathrm P} cos\left[g\Lambda_{{\cal I}_+}(x)\right], \ ~~~~~~ \
{\cal L}_O(x) = i{\mathrm P} sin\left[g\Lambda_{{\cal I}_+}(x)\right] \label{ksga}
\end{equation}
and the T-even (T-odd) part of the correlator consists of a series of even (odd) powers of $g$, each term being endowed with an even (odd) number of gluon legs. As a consequence, the zero order term is T-even, while the first order 
correction is T-odd. This confirms that no T-odd terms occur without interactions among partons, as claimed also by other authors (Brodsky {\it et al.}, 2002a,b, 2003; Collins, 2002). Gauge invariance of the correlator implies that these conclusions hold true in any axial gauge, such that condition (\ref{axi}) is fulfilled. From now on we shall work in such a type of gauge (Ji and Yuan, 2002; Belitsky {\it et al.}, 2003).
  
\subsection{Power Expansion of the Correlator}

We consider $\Phi_+$, which, as explained before, refers to DIS. As regards DY, the T-odd terms will change sign, as follows from the choice of the path - ${\cal I}_-$ instead of ${\cal I}_+$ - and from the first eq. (\ref{link0}) and 
from the second eq. (\ref{ksga}). We rewrite ${\cal L}(x)$ as
\begin{equation}
{\cal L}(x) = \sum_{n=0}^{\infty} (ig)^n\Lambda_n(x). \label{link01}
\end{equation} 
Here $\Lambda_0(x) = 1$, while for $n \geq 1$ one has, in the KS gauge,
\begin{equation}
\Lambda_n(x) = \int_{x_{1}}^{x_{2}}dz_1^{\mu_1} \int^{z_{1}}_{x_{1}} dz_2^{\mu_2} ... \int^{z_{n-1}}_{x_{1}}dz_n^{\mu_n} \left[{\bf A}_{\mu_n}(z_n) ... {\bf A}_{\mu_2}(z_2) {\bf A}_{\mu_1}(z_1)\right], \label{ll1}
\end{equation}
where the $z_i$ $\equiv$ $(\infty,z^+_i,{\bf z}_{i\perp})$, $i = 1, 2, ... n$, are points in the space-time along the line through $x_1$ and $x_2$. Substituting eq. (\ref{link01}) into eq. (\ref{corr1}), we have the following expansion 
of $\Phi$ in powers of $g$:
\begin{equation}
\Phi = \sum_{n=0}^{\infty} (ig)^n\Gamma_n, \label{exp}
\end{equation} 
with
\begin{equation}
(\Gamma_n)_{ij} = N\int\frac{d^4x}{(2\pi)^4} e^{ipx} 
\langle P,S|\bar{\psi}_j(0) \Lambda_n(x) \psi_i(x)|P,S\rangle. \label{coefn}
\end{equation}
As already noticed, $\Gamma_n$ is T-even for even $n$ and T-odd for odd $n$. 

Now we invoke the Politzer (1980) theorem, concerning EOM. This states that, if we consider the matrix element between two hadronic states of a given composite operator, constituted by quark and/or gluon fields, each such field fulfils EOM, no matters if the parton is off-shell and/or renormalized. We show in Appendix A that, owing to the Politzer theorem, the term $\Gamma_0$ fulfils the Dirac 
homogeneous equation, {\it i. e.},
\begin{equation}
(p\hspace{-0.45 em}/-m)\Gamma_0 = 0, \label{hom0}
\end{equation}
where $m$ is the quark rest mass. The corresponding Feynman-Cutkosky graph is represented in fig. 1. 
\begin{figure}[htb]
\begin{center}
\epsfig{file=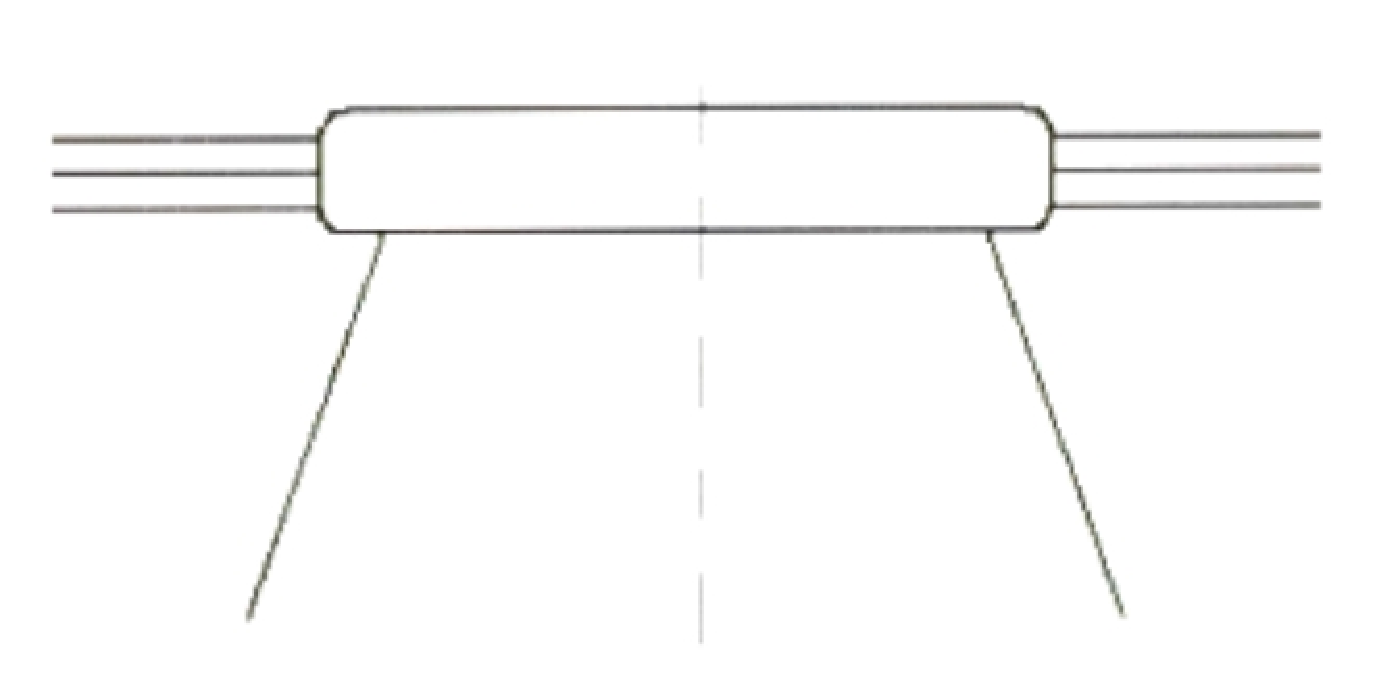, height=4.5cm, width=8.00cm}
\caption{Feynman-Cutkosky graph for zero order term of expansion (\ref{exp}).}
\end{center}
\label{fig:one}
\end{figure}
For $n \geq 1$ we have instead
\begin{equation}
(ig)^n\Gamma_n  = N\int d\Omega_n S^{\mu_1 ... \mu_n}\Phi^{(n)}_{\mu_1 ... \mu_n} (p, k_1, k_2 ... k_n). \label{pol}
\end{equation}
Here we have set
\begin{eqnarray}
d\Omega_n &=& \prod_{l=1}^n\frac{d^4k_l}{(2\pi)^4}, ~~~~~~~~~ \ ~~~~~~~~~~~~~~ \ ~~~~~~~~~~~~~~
\\
S^{\mu_1 ... \mu_n} &=& \frac{ig}{ p\hspace{-0.45 em}/-m+i\epsilon}i\gamma^{\mu_1} \frac{ig}
{ p\hspace{-0.45 em}/-\bar{k\hspace{-0.5 em}/}_1-m+i\epsilon}i\gamma^{\mu_2} ...\nonumber
\\
&\times&  \frac{ig}{ p\hspace{-0.45 em}/-\bar{k\hspace{-0.5 em}/}_{n-1}-m+i\epsilon}i\gamma^{\mu_n}, ~~~~~~~~~ \ ~~~~~~~~~ \label{propg}
\\
\bar{k}_l &=& \sum_{r=1}^l k_r. ~~~~~~~~~ \ ~~~~~~~~~ \ ~~~~~~~~~ \ ~~~~~~~~~ \ ~~~~~~~~~ \ 
\end{eqnarray}
The $k_r$ ($r = 1,2, ... n$) are the four-momenta of the $n$ gluons involved in the quark-gluon correlator $\Phi^{(n)}_{\mu_1 ... \mu_n}$. This is defined as
\begin{eqnarray}
&~& \left[\Phi^{(n)}_{\mu_1 ... \mu_n}(p, k_1, k_2 ... k_n)\right]_{ij} = N\int\frac{d^4x}{(2\pi)^4} e^{i(p-{\bar k}_n)x} \nonumber\\
&\times&  \langle P,S|\bar{\psi}_j(0){\mathrm P'}[{\bf B}_{\mu_n}(k_n) ... {\bf B}_{\mu_1}(k_1)] \psi_i(x)|P,S\rangle, \label{rec}
\end{eqnarray}
with
\begin{eqnarray}
{\bf B}_{\mu}(k) &=& {\hat {\bf A}}_{\mu}(k)+{\tilde {\bf A}}_{\mu}(k), ~~~~~~~~~ \ \\
{\hat {\bf A}}_{\mu}(k) &=& \int \frac{d^4z}{(2\pi)^4} {\bf A}_{\mu}(z) e^{ikz}, ~~~~~~~ \ \\
{\tilde {\bf A}}_{\mu}(k) &=& \delta(k^+) \displaystyle\lim_{M\to\infty}\int d\kappa e^{-i\kappa M}{\hat {\bf A}}_{\mu}(k^-,\kappa, {\bf k}_{\perp}). 
\end{eqnarray} 
Moreover the operator product ${\mathrm P'}$ is defined according to the following rules:

- any ${\hat {\bf A}}_{\mu}(k)$ is at the left of any ${\tilde {\bf A}}_{\mu}(k)$;

- the ${\tilde {\bf A}}_{\mu}(k)$ are ordered as ${\tilde {\bf A}}_{\mu_1}(k_1) {\tilde {\bf A}}_{\mu_2}(k_2) ... {\tilde {\bf A}}_{\mu_l}(k_l)$;

- the ${\hat {\bf A}}_{\mu}(k)$ are ordered as ${\hat {\bf A}}_{\mu_m}(k_m) ... {\hat {\bf A}}_{\mu_2}(k_2){\hat {\bf A}}_{\mu_1}(k_1)$.

Lastly the quark-gluon correlators $\Phi^{(n)}_{\mu_1 ... \mu_n}$ fulfil the following homogeneous equation:
\begin{equation}
(p\hspace{-0.45 em}/-\bar{k\hspace{-0.5 em}/}_n - m)\Phi^{(n)}_{\mu_1 ... \mu_n}(p, k_1, k_2 ... k_n) = 0. \label{homn}
\end{equation}

Each term of the expansion (\ref{exp}) - somewhat similar to the one obtained by Collins and Soper (1981, 1982) - may be interpreted as a Feynman-Cutkosky graph.  It corresponds to an interference term between the amplitude 
\begin{equation}
{\mathrm "nucleon} \rightarrow {\mathrm quark + spectator ~~ partons"}
\end{equation}
without any rescattering, and an analogous one, where $n$ gluons are exchanged between the active quark and the spectator partons. 
\begin{figure}[htb]
\begin{center}
\epsfig{file=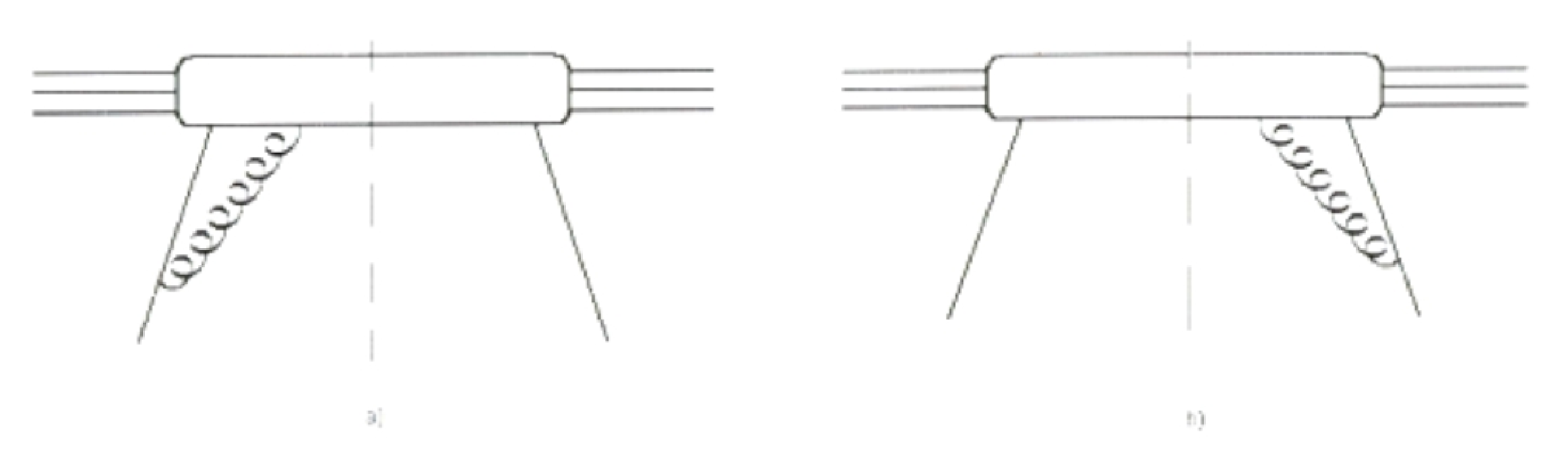, height=4.5cm, width=15.0cm}
\caption{Same as fig. 1 for first order correction in the coupling.}
\end{center}
\label{fig:two}
\end{figure}
In particular, the interference term is such that the gluons (for $n > 0$) are attached to the left quark leg, see figs. 2a and 3a. An important result, deduced at the end of Appendix A, is that such a term turns out to correspond to 
any interference term between two amplitudes, such that $k$  and $n-k$ gluons respectively are exchanged between the active quark and the spectator partons, with $0\leq k \leq n$. The situation is illustrated in figs. 2 and 3 for $n$ = 1 and 2. 

\begin{figure}[htb]
\begin{center}
\epsfig{file=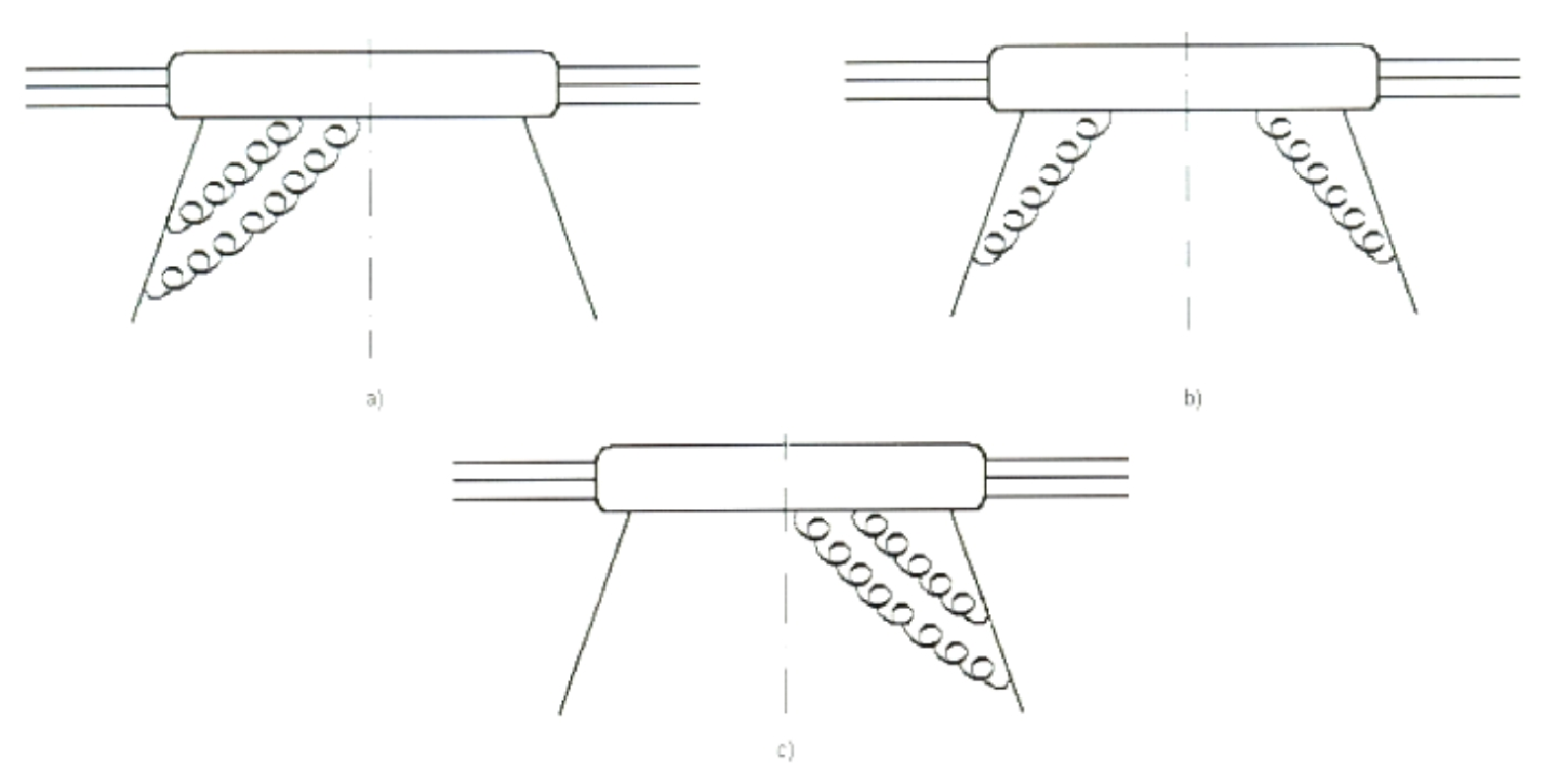, height=7.5cm}
\caption{Same as fig. 2 for second order correction.}
\end{center}
\label{fig:three}
\end{figure}

It is worth noting that a  radiation ordering similar to the one established here is found in semiinclusive processes at large $x$ (Catani {\it et al.}, 1991a) and in totally inclusive DIS at small $x$ (Catani {\it et al.}, 1991b). Moreover the terms (\ref{rec}) consist of quark-gluon-quark correlations, analogous to the one introduced by Efremov and Teryaev (1984) and by Qiu and Sterman (1991, 1992, 1998).

As a consequence of the Politzer theorem, formulae (\ref{exp}) to (\ref{rec}) hold for renormalized fields, provided we take into account the scale dependence of the coupling $g$, of the quark mass $m$ and of the correlators 
$\Phi^{(n)}_{\mu_1 ... \mu_n}(p, k_1, k_2 ... k_n)$ (Rogers, 2007). Moreover one has to observe that the four-momenta appearing in the propagators are highly off-shell: $p^2$ and $(p-\bar{k}_r)^2$ are of order $Q^2$ (Collins and Soper, 1982; Levelt and Mulders, 1994), because the uncertainty principle demands hard interactions to occur in a very limited space-time interval, corresponding to the condition
\begin{equation}
|p^2| \gg M^2.
\end{equation}
Therefore we have $p^2\approx 2p^+p^-$ and $p^+ = O(Q)$, whence 
\begin{equation}
|p^-| = O(Q) \label{heis}
\end{equation} 
and it follows that the coefficients $\Gamma_n$ are of order $Q^{-n}$, up to QCD corrections, consisting of terms of the type $g^{2k}(lnQ)^m$, with $k$ and $m$ integers and $k\geq m$ (Dokshitzer {\it et al.}, 1980). For the same reason, the coupling $g$, which appears in expansion (\ref{exp}), assumes small values, corresponding to short distances and times.

To summarize, we have found that the T-even and the T-odd correlators, given by eqs. (\ref{spl}), may be written as expansions in $gM/Q$, {\it i. e.},
\begin{equation}
\Phi_E(p) = \sum_{n=0}^{\infty} \left(\frac{igM}{Q}\right)^{2n}\bar{\Gamma}_{2n}, ~~~~~~~~ \ ~~~~~ \Phi_O(p) = 
\sum_{n=0}^{\infty} \left(\frac{igM}{Q}\right)^{2n+1}\bar{\Gamma}_{2n+1},\label{expp}
\end{equation}
where $\bar{\Gamma}_n = {\Gamma}_n Q^n/M^n$ has a relatively weak 
$Q$-dependence, as told above. Moreover, as already explained, $\Phi_O$ changes sign when involved in DY. Stated differently, T-odd terms present an odd number of quark propagators, see eq. (\ref{propg}) for odd $n$: in the limit of negligible quark mass, quark four-momenta in DIS are spacelike, whereas in DY they are timelike (Boer {\it et al.}, 2003b).

The first two terms of expansion (\ref{exp}) will be studied in detail in sects. 4 and 5 respectively.    

\section{Hadronic Tensor}

In the present section we refer indifferently to the hadronic tensor of one of the three processes introduced. To be precise, among these, only DY involves two correlators of the type illustrated in the last section, whereas SIDIS and $e^+e^-$ annihilation include respectively one and two fragmentation correlators. However, as we shall see in sect. 6, this object requires only minor modifications with respect to the correlator (\ref{corr1}).

If we substitute this correlator into the hadronic tensor (\ref{ht00}), this latter does not fulfil the requirement of electromagnetic gauge invariance: only the term of zero order in the coupling satisfies this condition. In order to get a complete gauge invariance at any order, we have to recall the interpretation given above of the correlator. For example, at first order in the coupling in SIDIS, we see that the "hard" scattering amplitude $q \gamma^* \rightarrow q' {\tilde g}$ - where we have denoted by $q$ and $q'$ the initial and final quark and by ${\tilde g}$ a gluon - consists not only of the graph of fig. 4a, encoded in the first order term of the correlator, but also of the one represented in fig. 4b, which interferes coherently with it. This guarantees electromagnetic gauge invariance for the first order graph (Berger and Brodsky, 1979). 
\begin{figure}[htb]
\begin{center}
\epsfig{file=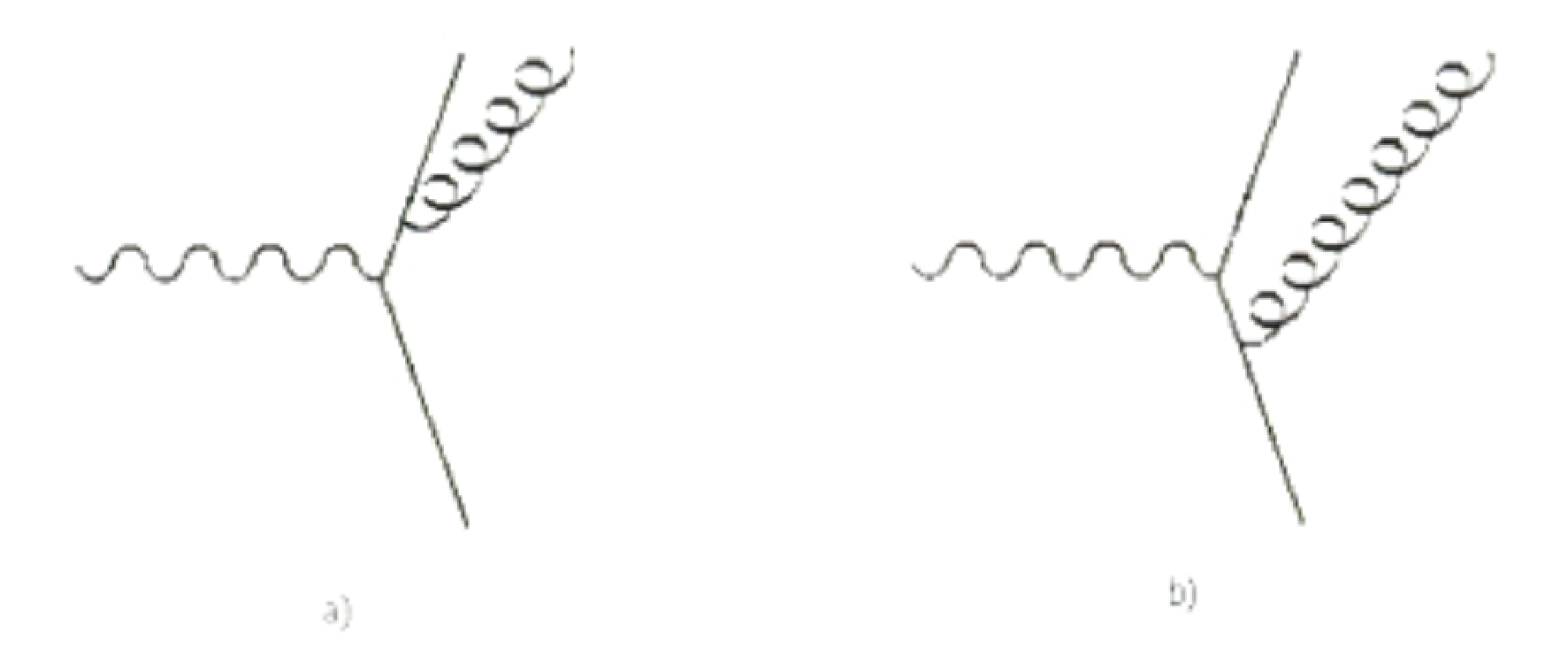, height=5.5cm, width=15.0cm}
\caption{Graphs for "hard" amplitudes interfering coherently, first order correction 
in the coupling.}
\end{center}
\label{fig:four}
\end{figure}
Furthermore, convoluting "hard" graphs with the "soft" factors, these two amplitudes give rise, among other objects, to asymmetric Feynman-Cutkosky graphs (fig. 5), related to interference terms. These are observables - necessarily gauge invariant - and therefore assume real values. This procedure, already suggested by LM, can be generalized to the three kinds of hadronic tensors considered in the present article, at any order in $g$, so as to obtain sets of graphs corresponding to observable, and therefore gauge invariant, quantities. We show how to construct them at any order $n$, corresponding to the overall number of gluons exchanged between active quarks and spectator partons. The procedure consists in the following steps, for a given $n$: 

- Consider the $n+1$ possible combinations of gluons occurring in the hadronic tensor (\ref{ht00}), say, $s$ for hadron 
$A$ and $n-s$ for hadron $B$, with $s$ = 0, 1 ... $n$.

- For a given $s ~~ (n-s)$, consider all possible correlators, according to the definition given in subsect. 2.2: as seen at the end of last section, these amount to $s+1 ~~ (n-s+1)$ correlators equal to $\Gamma_s ~~ (\Gamma_{n-s})$.

- Add each such correlator all those graphs whose "hard" parts interfere coherently with it, as shown in fig. 5. In practice, one has to do this for the correlator whose gluons are attached to the "left" quark leg and to multiply by 
the number of gluons of each correlator. 

\begin{figure}[htb]
\begin{center}
\epsfig{file=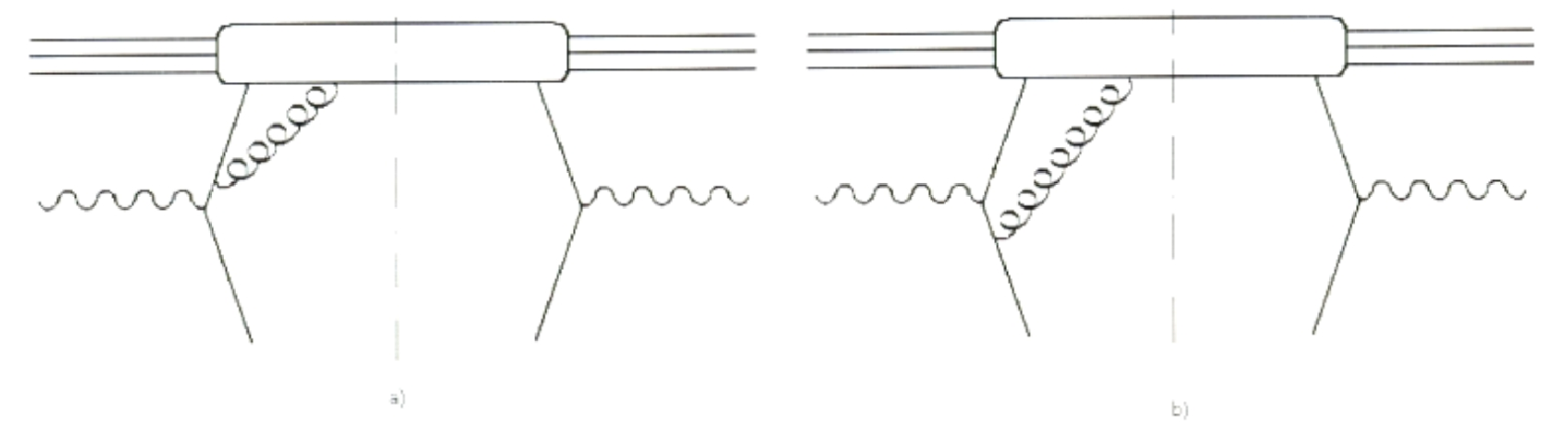, height=4.5cm, width=15.0cm}
\caption{Feynman-Cutkosky graphs corresponding to "hard" amplitudes of fig. 4. Also the complex conjugate graphs, which amount to mirror images of these two, contribute to first order corrections.}
\end{center}
\label{fig:five}
\end{figure}

Then we have, up to QCD corrections at each order of the expansion,
\begin{equation}
W_{\alpha\beta}(q) = \sum_{n=0}^{\infty} W_{\alpha\beta}^{(n)}(q), \label{tn}
\end{equation}
with
\begin{eqnarray}
W_{\alpha\beta}^{(n)} &=&  C \int \frac{d^4p}{(2\pi)^4}\int d\Omega_n \sum_{r=0}^n\sum_{s=0}^n Tr M_{\alpha\beta}^{(n)}, 
\label{tn00}
\\
M_{\alpha\beta}^{(n)} &=& \sum_{s=0}^n (s+1)(n-s+1) \left[{\tilde\Gamma}^{(s,0)}_{\alpha} \Phi_A^{s,0} 
{\tilde\Gamma}^{(n,s)}_{\beta} \Phi_B^{n,s}\right], \label{htg}
\\ 
{\tilde\Gamma}^{(l,r)}_{\rho} &=& \sum_{m=r}^l S_r^m \gamma_{\rho}S_m^l.
\end{eqnarray}
Here we have used the following shorthand notations:
\begin{equation}
S^m_r = S^{\mu_{r+1}, \mu_{r+2}, ..., \mu_m}, ~~~~~~ \ ~~~~~~~~~ \Phi^{n,s} = \Phi^{n,s}_{\mu_{s+1}, \mu_{s+2}, ..., \mu_n}. 
\end{equation}
$S^{\mu_{r+1}, \mu_{r+2}, ..., \mu_m}$ and $\Phi^{n,s}_{\mu_{s+1}, \mu_{s+2}, ..., \mu_n}$ are defined analogously to eqs. (\ref{propg}) and (\ref{rec}): the matrix product starts from $\mu_{r+1}$ and from $\mu_{s+1}$ respectively, rather than from $\mu_1$. In particular, $\Phi^{n,0}$ coincides with the definition (\ref{propg}). Last, we have set $S^r_r$ = 1.

For each term of expansion (\ref{tn}) we have to take into account three kinds of effects:

a) gluon radiation by scattered partons;

b) perturbative QCD corrections;

c) higher correlators, such that the active quarks exchange gluons with quark-antiquark pairs or gluon pairs or triplets belonging to spectator partons.

The first two effects may be calculated according to the algorithm suggested by Collins and Soper (1981, 1982). As to the contributions c), they can be included in the basic term of expansion (\ref{tn}), since they have the same (T-even or T-odd) behavior. Lastly we recall that, unless we integrate over some final transverse momentum [of the lepton pair in the case of DY, of a final hadron in SIDIS or $e^+e^-$ annihilation], the phase space of the final gluons emitted undergoes a restriction (Dokshitzer {\it et al.}, 1980), expressed by a 
doubly logarithmic form factor; this is more and more sizable at increasing energy,  resulting in the well-known Sudakov-like damping (Collins and Soper, 1981; Boer, 1999).  
   
\section{Zero order term: the QCD parton model} 

In this section and in the next one we elaborate the first two terms of the expansion of the hadronic tensor. To this end we define a suitable reference frame, such that the momentum ${\bf P }_B $ of the hadron $B$ has  an opposite direction to the momentum ${\bf P }_A$ of the hadron $A$, $|{\bf P }_A |$ and $|{\bf P }_B |$ are of order $Q$ and the $z$-axis is along ${\bf P }_A $. Moreover we focus on the hadronic tensor for DY process. However, as told at the beginning of the previous section, our results can be trivially extended to SIDIS and $e^+e^-$ annihilation; the main difference, concerning the fragmentation function, will be discussed in sect. 6.

Let us consider the hadronic tensor (\ref{tn00}) at zero order, {\it i. e.},
\begin{equation}
W_{\alpha\beta}^{(0)} = C \int \frac{d^4p}{(2\pi)^4} Tr \left[\gamma_{\alpha} \Gamma_0^A(p) \gamma_{\beta}\Gamma_0^B(p')\right]. \label{tn0}
\end{equation}
Here the $\Gamma_0$'s are given by eq. (\ref{coefn}), for $n$ = 0, and fulfil the homogeneous Dirac equation (\ref{hom0}). Incidentally, they are T-even and gauge invariant at zero order in $g$. Moreover $p'$ is defined by eq. (\ref{mom}). The tensor (\ref{tn0}), T-even itself, can be calculated, once we know the "soft" functions involved in the parameterizations of the correlators $\Gamma_0$'s. We show in Appendix B that 
\begin{equation}
\Gamma_0(p) = \frac{N}{4{\cal P}}(p\hspace{-0.45 em}/+m)\left[f_1(p)+ \gamma_5S\hspace{-0.65 em}/^q_{\parallel}g_{1L}(p)+\gamma_5S\hspace{-0.65 em}/^q_{\perp}h_{1T}(p)\right] 2p^+\delta(p^2-m^2). \label{cp0}
\end{equation}
Here $f_1(p)$, $g_{1L}(p)$ and $h_{1T}(p)$ are functions of the four-momentum $p$ of the active quark, which, in this case, is on shell: $p\equiv (E, {\bf p})$, with $E = \sqrt{m_q^2+{\bf p}^2}$. $S^q_{\parallel}$ and $S^q_{\perp}$ are the components of the quark PL vector, respectively parallel and perpendicular to the hadron momentum. Moreover we have set
\begin{equation}
{\cal P} =\frac{1}{\sqrt{2}}p\cdot n_-, \label{cns}
\end{equation}
having defined the dimensionless, light-like four-vectors $n_\pm$ in such a way that  
\begin{equation}
n_+\cdot n_- = 1 \label{cnt}
\end{equation}
and such that their spatial components are along (+) or opposite (-) to the hadron momentum. It is important to notice that, if integrated over $p^-$, the expression obtained for the zero order correlator turns out to be proportional to the density matrix of a  quark confined in a finite volume, but free of interactions with other partons (Di Salvo, 2007b). Therefore we fix the normalization constant $N$ so as to obtain, after integration, just the density matrix  {\it i. e.}, 
\begin{equation}
N = 2{\cal P}. \label{cnu}
\end{equation}
Lastly, it is convenient to express $S^q_{\parallel}$ and $S^q_{\perp}$ in terms of the components of the PL vector of the hadron. As shown in Appendix B, one has
\begin{equation}
S^q_{\parallel} = \lambda \left(\frac{\bar{p}}{m}\right)-\bar{\eta}_{\perp} + O(\bar{\eta}^2_{\perp}), ~~~~~~~~~~~ \ 
~~~~~~ S^q_{\perp} =  S_{\perp}+\bar{\lambda}_{\perp}\frac{\bar{p}}{m}+ O(\bar{\eta}^2_{\perp}).  \label{qsp1}
\end{equation}
Here 
\begin{eqnarray}
\lambda &=& -S\cdot \frac{n_+ + n_-}{\sqrt{2}}, ~~~~~ \ ~~~~~~ \ ~~~ ~~~~~ \ ~~~~~ \
S_{\bot} = S -\lambda\frac{n_+ + n_-}{\sqrt{2}},
\\
\bar{p}&\equiv& (|{\bf p}|, E\hat{\bf p}), ~~~~~ \ ~~~~~ {\hat{\bf p}} = {\bf p}/|{\bf p}|, ~~~~~ \ ~~~ \bar{\eta}_{\perp} = p_{\perp}/{\cal P},
\\
\bar{\lambda}_{\perp} &=& -S\cdot\bar{\eta}_{\perp}, ~~~~~ \ ~~~~~ \ ~~~~~~ \ ~~~~~ \  ~~~~~ \ ~~~~~ \ p_{\perp}\equiv(0,0,{\bf p}_{\perp})
\end{eqnarray}
 and ${\bf p}_{\perp}$ is the transverse momentum of the active quark with respect to the hadron momentum. 

Equation (\ref{cp0}) has important consequences on TMD T-even functions, as we are going to illustrate in the two next subsections. To this end we compare that equation with the naive parameterization of the TMD correlator in terms of 
Dirac components, without introducing any dynamic conditions (Mulders and Tangerman, 1996; Boer {\it et al.}, 2000; Goeke {\it et al.}, 2005). We give such a parameterization in Appendix C, up to and including twist-3 terms. The twist-2, T-even sector, which we study in subsect. 4.1, corresponds 
to quark distribution functions which survive when interactions with gluons are turned off. As regards the twist-3 functions, we distinguish among the T-even, the T-odd and the "hybrid" ones, these lasts deriving contributions both 
from T-even and T-odd terms.   

\subsection{Twist-2, T-even Correlator}

If quark-gluon interactions are neglected, the correlator includes just twist-2, T-even terms. We show in Appendix C that it can be parameterized as
\begin{eqnarray}
\Phi^f_E &=& \frac{\cal P}{\sqrt{2}}  \{f_1n\hspace{-0.5 em}/_+ + (\lambda 
g_{1L}+\lambda_{\perp}g_{1T})\gamma_5n\hspace{-0.5 em}/_+ + \frac{1}{2}h_{1T}\gamma_5[S\hspace{-0.65 
em}/_{\perp},n\hspace{-0.5 em}/_+]\nonumber
\\ 
&+& \frac{1}{2}(\lambda h^{\perp}_{1L}+\lambda_{\perp} h^{\perp}_{1T}) \gamma_5 [\eta\hspace{-0.5 em}/_{\perp},n\hspace{-0.5 em}/_+]\}2p^+\delta(p^2-m^2). \label{par02}
\end{eqnarray}
Here we have adopted the usual notations for the non-perturbative functions (Kotzinian, 1995; Tangerman and Mulders, 1995); the indices $f$ and $E$ of $\Phi$ denote respectively the feature of "free" and "T-even". The Dirac operators considered are purely T-even, as can be checked; moreover
\begin{equation}
\eta_{\perp} = p_{\perp}/{\mu_0}, ~~~~~~ \ ~~~~~ \lambda_{\perp} = -S\cdot \eta_{\perp} \label{etap}
\end{equation}
and $\mu_0$ is an undetermined energy scale, introduced for dimensional reasons, in such a way that all functions embodied in the parameterization of $\Phi$ have the dimensions of a probability density. This scale (Kotzinian, 1995)
determines the normalization of the functions which depend on $\eta_{\perp}$. In particular, as is well-known, the 6 twist-2 functions, which appear in the parameterization (\ref{par02}), are interpreted as TMD probability densities: 
$f_1$ is the unpolarized quark density, $g_{1L}$ the longitudinally polarized density in a longitudinally polarized 
(spin 1/2) hadron, $g_{1T}$ the longitudinally polarized density in a transversely polarized hadron, $h^{\perp}_{1L}$ 
the transversity in a longitudinally polarized hadron and 
\begin{equation}
h'_{1T} = h_{1T}+|\eta_{\perp}^2|h^{\perp}_{1T} \label{TMDt}
\end{equation}
is the TMD transversity in a transversely polarized 
hadron\footnote{The function $h^{\perp}_{1T}$ is known as "pretzelosity" (Avakian {\it et al.}, 2008b).}.

Now we compare the parameterization (\ref{par02}) with the correlator (\ref{cp0}). To this end we consider projections of both matrices over the various Dirac components, {\it i. e.}, for a given Dirac operator $\Gamma$,
\begin{equation}
\Phi^{\Gamma} = \frac{1}{2}Tr{\Gamma\Phi},
\end{equation}
taking into account eqs. (\ref{qsp1}) wherever necessary.

First of all, $\Gamma$ = $\gamma_5\gamma^+$ and $\gamma_5\gamma^+\gamma_i$ ($i$ = 1, 2) yield, approximately in the limit of $m$ = 0,
\begin{equation}
h^{\perp}_{1L} \approx -\frac{\mu_0}{\cal P}g_{1L}, ~~~~~ \ ~~~~ g_{1T} \approx  \frac{\mu_0}{\cal P}h_{1T}, ~~~~~ \ 
~~~~ h_{1T}^{\perp}\approx \frac{\mu_0^2}{{\cal P}^2}h_{1T}.\label{ff1}
\end{equation}
These relations hold up to terms of order $(gM/Q)^2$, since, as we have seen, the T-even Dirac components of $\Phi$ derive contributions only from even powers of $gM/Q$. Moreover, the Politzer theorem implies that the relations are not modified by renormalization effects, and therefore hold also taking into account QCD evolution.  

In order to determine $\mu_0$, we observe that the functions involved in both sides of eqs. (\ref{ff1}) are independent of ${\cal P}$. Therefore we must set $\mu_0 = C_0 {\cal P}$, $C_0$ being a dimensionless numerical constant, 
independent of momentum. But since these functions are quark densities, they should be normalized adequately, setting $C_0$ = 1. Then, neglecting the quark mass,
\begin{equation}
\mu_0 = {\cal P} = \frac{1}{\sqrt{2}}p\cdot n_-. \label{mu0}
\end{equation}
This result differs from the treatments of previous authors (Mulders and Tangerman, 1996; Goeke {\it et al.}, 2005), who assume $\mu_0 = M$. Some mismatches have been shown, as consequences of this choice (Bacchetta {\it et al.}, 2008); these could be eliminated by taking into account result 
(\ref{mu0}). 

By comparing CLAS (Avakian {\it et al.}, 2005) and HERMES (Airapetian {\it et al.}, 2005b) results, at not too high values of $Q^2$ (1.5 to 3 $GeV$) the first 
relation (\ref{ff1}), together with eq. (\ref{mu0}), is verified for $x < 0.35$ (Di Salvo, 2007b), discrepancies at larger $x$ being attributed to higher twist contributions.   

\subsection{Twist-3, "Hybrid" Correlator}

Now we consider a sector of the correlator which, as explained above in this section, has both T-even and T-odd contributions. In particular, here we focus on that part of "hybrid" correlator which comes from the so-called "kinematic" twist-3 terms. In Appendix C we find, according to the usual notations (Mulders and Tangerman, 1996; Goeke {\it et al.}, 2005), 
\begin{eqnarray}
&~&\Phi_H^f =  \{\frac{1}{2}(f^{\perp}+\lambda g^{\perp}_L\gamma_5 + \lambda_{\perp} g^{\perp}_T\gamma_5)p\hspace{-0.45 
em}/_{\perp} + \frac{1}{4}\lambda_{\perp} h_T^{\perp} \gamma_5 [S\hspace{-0.65 em}/_{\perp}, p\hspace{-0.5 em}/_{\perp}] \nonumber
\\
&+& \frac{1}{2}x M \left(e + g'_T\gamma_5 S\hspace{-0.65 em}/_{\perp} + \frac{1}{2}(\lambda h_L  + \lambda_{\perp} h_T) 
\gamma_5 [n\hspace{-0.5 em}/_-,n\hspace{-0.5 em}/_+]\right)\}2p^+\delta(p^2-m^2). \label{par03}
\end{eqnarray}
Comparing the operator (\ref{par03}) with the correlator (\ref{cp0}), and considering, in particular, the projections of over $\Gamma$ = $\gamma_i$ ($i$ = 1, 2) of such operators, yields the approximate relation
\begin{equation}
f^{\perp} \approx f_1,\label{ff2}
\end{equation}
which corresponds to the Cahn (1978, 1989) effect and is approximately verified for sufficiently large $Q^2$ and small $x$ (Anselmino {\it et al.}, 2007). Also this equation, like eqs. (\ref{ff1}), survives QCD evolution. As we shall see in the next section, eq. (\ref{ff2}) holds up to terms of order $gM/Q$, since $f^{\perp}$ derives also T-odd contributions from one-gluon exchange. 

The projections of the same operators over $\Gamma$ = $\gamma_5\gamma_i$ ($i$ = 1, 2) yield (after integration over ${\bf p}_{\perp}$)
\begin{equation}
g_T(x) \approx \frac{m}{xM} h_1(x). \label{ff3}
\end{equation}
Here
\begin{equation}
g_T(x) = \int d^2p_{\bot} g'_T(x, {\bf p}_{\perp}^2) \label{fg0}
\end{equation}
and
\begin{equation} 
h_1(x) = \int d^2 p_{\perp} \left[h_{1T}(x,{\bf p}_{\perp}^2)+
 |\eta_{\perp}^2| h_{1T}^{\perp}(x,{\bf p}_{\perp}^2)\right]. \label{trr1} 
\end{equation}
This last equation has been obtained from eq. (\ref{TMDt}). The contribution of the QCD parton model to $g_T(x)$ is very small: $m$ is negligible for $u$- and $d$-quarks, while for $s$-quarks $h_1$ is presumably small, because the sea is produced mainly by annihilation of gluons, whose transversity is zero in a nucleon. Therefore the contribution of quark-gluon interactions, neglected in the approximation considered, becomes prevalent in this case, as well as for $\Gamma$ = $1$ and $\gamma_5\gamma_+\gamma_-$, corresponding respectively to the functions $e$ and $h_{L}$ in eq. (\ref{par03}). The effect of such interactions will be discussed in sect. 5. 

\subsection{Remarks}

To conclude this section, we sketch some consequences of our theoretical results.

A) In the expression (\ref{TMDt}) or (\ref{trr1}) for transversity, the second term is due to a relativistic effect. To illustrate this, consider a transversely polarized hadron. The longitudinal polarization of the quark, due in this case to the transverse momentum, is magnified by the boost from the quark rest frame. This additional polarization, along the quark momentum, has again a transverse component with respect to the nucleon momentum.

B) Eq. (\ref{trr1}), together with the last two eqs. (\ref{ff1}), suggests a method for determining approximately the nucleon transversity. Indeed, $g_{1T}$ can be conveniently extracted from double spin asymmetry (Kotzinian and Mulders, 1996; Di Salvo, 2002, 2003) in SIDIS with a transversely polarized target. This asymmetry is expressed as a convolution of the unknown function 
with the usual, well-known fragmentation function of the pion. Therefore the method appears complementary to the one usually proposed (Airapetian {\it et al.}, 2000; Anselmino {\it et al.}, 2007), based on the Collins (1993) effect
in single spin SIDIS asymmetry; in this latter case one is faced with the convolutive product of $h_{1T}$ with the Collins function, which is poorly known (Efremov {\it et al.}, 2006a,b). 

C) Eq. (\ref{ff3}) establishes a relation between transversity and transverse spin. Indeed, the two quantities are related to each other. But, unlike transversity, the transverse spin operator is chiral even and does not commute with the free hamiltonian of a quark (Jaffe and Ji, 1991a): in QCD parton model it is proportional to the quark rest mass, which causes chirality flip. 

D) We note that $g_{1T}$, $h^{\perp}_{1L}$ and $h^{\perp}_{1T}$ are associated with "twist-2" Dirac operators (Jaffe and Ji, 1991a, 1992), and yet, in our treatment, they are multiplied by inverse powers of $Q$, as results from eqs. (\ref{par02}) and (\ref{mu0}): $Q^{-1}$ for the first two functions, $Q^{-2}$ for the third one. This would be unacceptable for common distribution functions; but, when transverse momentum is involved, also the orbital angular momentum plays a role. To illustrate this point, we recall that the quark distribution functions may be regarded as the absorptive parts of $u$-channel quark-hadron amplitudes (Soffer, 1995). For example, $g_{1T}$ corresponds to an amplitude of the type $\langle++|-+\rangle$, denoting by $|\Lambda\lambda\rangle$ a state in which the nucleon and quark helicities are, respectively, $\Lambda$ and $\lambda$. The amplitudes corresponding to the functions in question involve a change $\Delta L = 1$ (for $g_{1T}$ and $h^{\perp}_{1L}$) or $\Delta L = 2$ (for $h^{\perp}_{1T}$) in the orbital angular momentum; therefore they are of the type 
\begin{equation}
{\cal A} = A (sin \theta)^{\Delta L}, \label{ampl}
\end{equation}
where $\theta$ = $arcsin |{\bf p}_\bot|/|{\bf p}|$ is the angle between the nucleon momentum and the quark momentum, while $A$ is weakly energy dependent. But $|{\bf p}|$ is of order $Q$ and $|{\bf p}_\bot|$ of order $M$. Therefore eq. (\ref{ampl}) reproduces the $Q$-dependence of the coefficients relative to the above mentioned functions\footnote{This observation is the fruit of a stimulating discussion  with Nello Paver.}.

\section{First Order Correction}

The first order correction in $g$ of the hadronic tensor reads [see eqs. (\ref{tn00}) and (\ref{htg})]
\begin{equation}
W_{\alpha\beta}^{(1)} = -2g C \int \frac{d^4p}{(2\pi)^4} \int\frac{d^4k}{(2\pi)^4} Tr N_{\alpha\beta}. \label{tn1}
\end{equation}
Here we have set
\begin{equation}
N_{\alpha\beta} = 2 [h_{\alpha}^{\mu}(p,p',k)\Phi_{A\mu}^{(1)}(p,k) 
\gamma_{\beta}\Gamma^B_0(p')+\gamma_{\alpha}\Gamma_0^A(p) h_{\beta}^{\mu}(p',p,k) \Phi_{B\mu}^{(1)}(p',k)] \label{mab}
\end{equation}
and
\begin{equation}
h_{\alpha}^{\mu}(p,p',k) = \gamma_{\alpha}\frac{1}{p\hspace{-0.45 em}/-m+i\epsilon}\gamma^{\mu} + \gamma^{\mu}\frac{1}{p\hspace{-0.45 em}/'- 
k\hspace{-0.5 em}/-m+i\epsilon}\gamma_{\alpha}. \label{hmn} 
\end{equation}
Furthermore the $\Phi_{\mu}^{(1)}$'s are given by eq. (\ref{rec}) for $n$ = 1 and fulfil the homogeneous Dirac equation
\begin{equation}
(p\hspace{-0.45 em}/- k\hspace{-0.5 em}/-m)\Phi_{\mu}^{(1)}(p,k) = 0. 
\end{equation}
Therefore, in the gauge adopted, this function is parameterized as
\begin{equation}
\Phi^{(1)}_{\mu}(p,k) = \Psi_{\mu}(p, k)\delta\left(p_1^--\frac{{m}^2+{\bf p}^2_{1\perp}}{2p_1^+}\right). \label{fst}
\end{equation} 
Here we have set
\begin{equation}
p_1 = p-k, ~~~~~~~~ {\mathrm with} ~~~~~~~~ p_1^2 = m^2 \label{mmt}
\end{equation} 
and
\begin{equation}
\Psi_{\mu}(p,k) \approx \frac{1}{2}(p\hspace{-0.45 em}/_1+m) {\hat L} [{\cal C}_{\mu}+\Delta{\cal C}_{\mu}\gamma_5S\hspace{-0.65 em}/^q_{\|}+\Delta_T
{\cal C}_{\mu}\gamma_5S\hspace{-0.65 em}/^q_{\bot}+\Delta_T
{\cal C}'_{\mu}\gamma_5{\bar S}\hspace{-0.65 em}/_{\bot}]. \label{psi1}
\end{equation}
This is a consequence of the Politzer theorem, as shown in Appendix B. The 
quantities ${\cal C}_{\mu}$, $\Delta{\cal C}_{\mu}$, $\Delta_T{\cal C}_{\mu}$ and $\Delta_T{\cal C}'_{\mu}$ are correlation functions of $p$ and $k$. In particular, we have (see subsect. B.2)
\begin{eqnarray}
{\cal C}_{\mu} &=& p_{1\perp\mu}{\cal C}_1 + \epsilon_{\mu\nu\rho\sigma}
n_-^{\nu} ({\cal C}_2 \lambda n_+^{\rho}p_{1\perp}^{\sigma} +{\cal C}_3 MS_{\bot}^{\rho}n_+^{\sigma}),  \label{cmi}
\\
\Delta{\cal C}_{\mu} &=& \Delta{\cal C} p_{1\perp\mu}, \label{dcmi}
\\
\Delta_T{\cal C}_{\mu} &=& \Delta_T{\cal C} p_{1\perp\mu}, \label{dtcmi} 
\\
\Delta_T{\cal C}'_{\mu} &=& \Delta_T{\cal C}' p_{1\perp\mu}. \label{bm}
\end{eqnarray}
Here the ${\cal C}_i$ ($i$ = 1,2,3) are unpolarized. $\Delta{\cal C}$ ($\Delta_T{\cal C}$) is a longitudinally (transversely) polarized function in a longitudinally (transversely) polarized nucleon. $\Delta_T{\cal C}'$ is a transversely polarized correlation function in an unpolarized nucleon: it is connected to quark-gluon interaction, for example, to a spin-orbit coupling (Brodsky {\it et al.}, 2002a,b, 2003).
 
Last, we have set in eq. (\ref{psi1})
\begin{eqnarray}
\sqrt{|p^2_{\bot}|}{\bar S}_{\bot \alpha} &=& \epsilon_{\alpha\beta\rho\sigma}n_+^{\beta}n_-^{\rho} p_{\bot}^{\sigma}, 
~~~~~ \ ~~~~~ \ ~~~
\\ 
{\hat L} &=& \sqrt{\frac{{\cal P}}{{\cal P}_1}}\left[cosh\varphi+\gamma_0\gamma_3 a\frac{sinh\varphi}{2\varphi}\right]. 
\label{lor1}
\end{eqnarray}
Here ${\cal P}_1 = p_1^+/\sqrt{2}$, while $\varphi$ and $a$ are defined in Appendix B. 

\subsection{Approximate Factorization}

The second term of eq. (\ref{hmn}) is not factorizable, in agreement with the observations of various authors (Brodsky {\it et al.}, 2002a,b, 2003; Peign\'e, 2002; Collins and Qiu, 2007), who have shown failures of universality (Peign\'e, 2002; Collins and Qiu, 2007) at large tranverse momentum. 
However, for sufficiently large $Q$, and adopting an axial gauge, this term is negligibly small (Berger and Brodsky, 1979) in comparison with the first one, which instead is factorizable. In fact, the gluon corresponding to the first term has a smaller offshellness than the one involved in the second term. This approximation is especially acceptable, even for relatively 
small $Q$, provided we limit ourselves to small transverse momenta (Collins, 2002) of the initial hadrons with respect to the direction of the momentum of the virtual photon in the center of mass of the DY pair. However, as already explained in sect. 2, also in the case when factorization is approximately satisfied, the T-odd distribution functions change sign 
from SIDIS to DY. We shall illustrate phenomenological implications of this change of sign in sect. 7.

In this approximation the tensor (\ref{tn1}) amounts to
\begin{equation}
W_{\alpha\beta}^{(1)} = -4g C \int \frac{d^4p}{(2\pi)^4} \gamma_{\alpha} [\Gamma^{A}_1(p)\gamma_{\beta}\Gamma^B_0(p')+\Gamma_0^A(p)\gamma_{\beta}
\Gamma^{B}_1(p')], \label{tn2a}
\end{equation}
where
\begin{equation}
\Gamma_1(p) =   \frac{1}{p\hspace{-0.45 em}/-m+i\epsilon} \gamma^{\mu}\int\frac{d^4k}{(2\pi)^4}\Phi^{(1)}_{\mu}(p,k) 
\label{tn2b}
\end{equation}
and $\Gamma_0$ is given by eq. (\ref{cp0}).
Then the tensor $W_{\alpha\beta}^{(1)}$ assumes a form similar to $W_{\alpha\beta}^{(0)}$, giving rise to an approximate (Brodsky {\it et al.}, 2002a) factorization of T-odd functions. Our conclusion is quite analogous to the one drawn by Collins (2002) and presents some similarity with the Qiu-Sterman (1991) assumption about the quark-gluon-quark correlation functions.
In particular, as regards the factors $\Gamma_1(p)$, defined by eq. (\ref{tn2b}), we have to take into account eqs. (\ref{fst}) to (\ref{bm}), together with eqs. (\ref{qsp1}). These induce for $\Gamma_1$ the following parameterization, at twist-3 approximation: 
\begin{eqnarray}
\Gamma_1(p) &\approx& \frac{2p^+}{\pi(p^2-m^2+i\epsilon)}\frac{1}{2}\gamma_-\gamma_+[p\hspace{-0.45 em}/_{\bot}f_o^{\bot} + \gamma^i \epsilon_{i\nu\sigma\rho} n^{\nu}_- (\lambda p_{\bot}^{\sigma}n_+^{\rho} f_L^{\bot} \nonumber
\\
&+& M n_+^{\sigma}S_{\bot}^{\rho}g'_{T,o}) + \gamma_5 S\hspace{-0.65 em}/_{\bot} p\hspace{-0.45 em}/_{\bot} h_{T,o} +
\gamma_5 {\bar S}\hspace{-0.65 em}/_{\bot} p\hspace{-0.45 em}/_{\bot} h' +\lambda\gamma_5p\hspace{-0.45 em}/_{\bot} g^{\bot}_{L,o}].  \label{pargam}
\end{eqnarray}
Here we have defined 
\begin{eqnarray}
f^{\bot}_o &=& -\int d{\tilde \Omega} {\cal C}_1, ~~~~ \ ~~~ f^{\bot}_L = \int 
d{\tilde\Omega}({\cal C}_2+r\Delta{\cal C}), ~~~~ \ ~~~  
g'_{T,o} = \int d{\tilde \Omega} {\cal C}_3, \label{fgo}
\\
h_{T,o} &=&  -\int d{\tilde \Omega} \Delta_T{\cal C}, ~~~~ \ ~~~
h' = \int d{\tilde \Omega} \Delta_T{\cal C}',  ~~~~ \ ~~~  g^{\bot}_{L,o} = \int d{\tilde \Omega} {\cal C}_2. ~~~~ \ ~~~ \label{hg}
\end{eqnarray}
Moreover
\begin{eqnarray}  
d{\tilde \Omega} &=& \pi\frac{d^3{\tilde k}}{(2\pi)^4}\frac{p^-p_1^+}{2p^+}L^{(-)}, ~~~~~ \ ~~~~~ \ ~~~~~ \ ~~~~~ r = 
\frac{k^-{\bar p}_0^+}{p_1^+p^-}\frac{L^{(+)}}{L^{(-)}},
\\
L^{(\pm)} &=& \frac{{\cal P}}{{\cal P}_1}\left[cosh\varphi\pm a\frac{sinh\varphi}{2\varphi}\right] ~~~~~ {\mathrm and} 
~~~~~ d^3{\tilde k} = 2p_1^+d^4p_1\delta{(p_1^2-m^2)}.
\end{eqnarray}
Lastly $p_1$ is defined by eqs. (\ref{mmt}) and
\begin{equation}
{\bar p}_0\equiv\left(|{\bf p}|,{\bf p}\frac{\sqrt{{\bf p}^2+m^2}}{|{\bf p}|}\right).
\end{equation}

The notations for the functions are somewhat similar to those introduced by Mulders and Tangerman (1996) and Goeke {\it et al.} (2005). The suffix $"o"$ in 
$f^{\bot}_o$, $g_{T,o}$, $g^{\bot}_{L,o}$ and $h_{T,o}$ denotes T-odd contribution to these three functions. They have T-even counterparts, as
explained in  sect. 4, eq. (\ref{par03}), where we introduced "hybrid" functions. The T-odd functions are normalized coherently with their T-even counterparts, as can be seen from the factor in front of $\Gamma_1$, eq. (\ref{pargam}): indeed, considering the case of an approximately on-shell quark, we have
\begin{equation}
[\pi(p^2-m^2+i\epsilon)]^{-1} \to -i\delta(p^2-m^2). \label{propd}
\end{equation}
Furthermore the $(-i)$-factor in (\ref{propd}) is compensated by the $i-$factor present in the term with $n = 1$ in expansion (\ref{exp}), but absent in the term with $n = 0$; therefore also the phase of the T-odd functions is in agreement with the one of the T-even counterparts. It follows from such observations that the factor (\ref{propd}) in expression (\ref{pargam}) automatically fixes also the normalization and the phase of the remaining functions included in $\Gamma_1$.

Last, as already noticed in connection with correlation functions, the function $h'$ describes a quark transverse polarization induced by quark-gluon interactions: this polarization, present also in spinless or unpolarized hadrons, is somewhat similar to the Boer-Mulders (1998) function, although it is twist-3 and not twist-2.
  
\subsection{Twist-3, T-odd correlator}

As explained in the previous subsection, $\Gamma_1(p)$, eq. (\ref{pargam}), yields, in the approximation discussed above, the contribution to the quark correlator of quark-gluon interactions, at $Q^{-1}$ approximation. We compare this expression with the purely kinematic parameterization of the twist-3, interaction dependent correlator, as given in appendix C. In this way we obtain several approximate relations among the "soft" functions involved in that parameterization. This last reads
\begin{equation}
\Phi^{i} = \Phi^{i}_H + \Phi^{i}_O.\label{idcor}
\end{equation}
Here $\Phi^i_H$ is obtained from eq. (\ref{par03}), by substituting $\delta(p^2-m^2)$ by $[\pi(p^2-m^2+i\epsilon)]^{-1}$, according to the rule just stated at the end of subsect. 5.1. On the other hand, from Appendix C we get
\begin{eqnarray}
\Phi^i_O &=& \frac{2p^+}{\pi(p^2-m^2+i\epsilon)}
\{\epsilon_{ij}S_{\bot}^i(p_{\bot}^j e_T^{\bot} + M \gamma^j f_T)+ 
\epsilon_{ij}{\bar S}_{\bot}^{i} p_{\bot}^j e_T^{'\bot}+\gamma_5 (xM e_L\lambda \nonumber  
\\
&+& e_T p_{\bot}\cdot S_{\bot} + e'_T p_{\bot}\cdot {\bar S}_{\bot})  + \epsilon_{ij} \gamma_i p_{\bot}^j 
(f_L^{\bot}\lambda + f_T^{\bot}\lambda_{\bot} + \gamma_5 g^{\bot}) \nonumber 
\\  
&+&\gamma_5 p\hspace{-0.45 em}/_{\bot} {\bar S}\hspace{-0.65 em}/_{\bot}h' +\frac{1}{2}\gamma_5 [\gamma_+,\gamma_-]p_{\bot}\cdot 
{\bar S}_{\bot}h^{'\bot}\}.
\end{eqnarray}
Comparison between parameterization (\ref{idcor}) and result (\ref{pargam}), component by component, yields the following approximate relations:
\begin{eqnarray}
g^{\bot} &\approx& f_o^{\bot}, ~~~~~~~~ f^{\bot}_{L} \approx g^{\bot}_{L,o}, ~~  ~~ f_T\approx g'_{T,o}, \label{fg}
\\
e_T &\approx& -e_T^{\bot} ~~~ \approx ~~ h_{T,o}^{\bot} ~~ \approx  ~~ h_{T,o}, \label{htt}
\\
e'_T &\approx& -e_T^{'\bot} ~~~ \approx ~~ h^{'\bot} ~~ \approx  ~~ h', \label{htp}
\\
e_L &\approx& f_T^{\bot} ~~ \approx ~~ g_{T,o}^{\bot} ~~ \approx ~~ e_o ~~ \approx ~~ h_{L,o} \approx 0. \label{fz}
\end{eqnarray}
Also these equations survive QCD evolution, like eqs. (\ref{ff1}) and (\ref{ff2}). Aside from that, it is important to notice that the second eq. 
(\ref{fg}) implies, together with the second eq. (\ref{fgo}) and with the third eq. (\ref{hg}), 

a) that $\Delta{\cal C}$ = 0;

   b) that $\Gamma_1$ includes 5 independent functions in all.

\subsection{Remarks}

~~~ A) Some of the functions, which appear in the equalities (\ref{fg}) to (\ref{htp}), are longitudinally ($g^{\bot}$, $g^{\bot}_{L,o}$) or transversely ($h^{'\bot}$, $h'$) polarized in an unpolarized nucleon. Conversely, other functions are unpolarized in a longitudinally ($f^{\bot}_{L}$) or transversely ($f_T$ and the "$e$"-functions) polarized nucleon\footnote{$f_T$ is known as the Sivers (1990, 1991) function}. This is a consequence of the spin-orbit coupling (Brodsky {\it et al.}, 2002a) in gluon-quark interactions.
Furthermore, unlike previous authors (Boer and Mulders, 1998; Boer {\it et al.}, 2000; Goeke {\it et al.}, 2005), we find that such functions are are associated to the same inverse power of $Q$, independent of the kind of polarization
(longitudinal or transverse) of the quark or of the nucleon.

B) Among eqs. (\ref{fg}) to (\ref{fz}), those which concern only T-odd functions hold up to terms of order $(gM/Q)^2$. On the contrary, those which involve "hybrid" functions - including eq. (\ref{ff2}) - hold up to terms of order $gM/Q$. Analogous approximate relations of this latter type have been found by Avakian {\it et al.} (2008a) and by Efremov {\it et al.} (2009).

C) By integrating the correlator (\ref{pargam}) over the transverse momentum of the quark, we obtain interesting results as regards twist-3 common functions. First of all, the fourth eq. (\ref{fz}) implies that $e(x)$ derives just 
T-even contributions, and therefore, apart from the (negligible) term illustrated in the previous section, it is essentially of order $(gM/Q)^2$. On the contrary, the main contributions to $g_T$ and $h_L$ are of order $gM/Q$ and are T-odd; therefore they change sign according as to whether they are involved in DIS or DY reaction. These last predictions could be tested by confronting the DIS double spin asymmetry (Anthony {\it et al.}, 1996a,b, 2003) with the DY one (Di Salvo, 2001; Soffer and Taxil, 1980). In the case of DY one has to integrate over the transverse momentum of the virtual photon; moreover, if possible, it may be more promising to detect $\tau^+\tau^-$ pairs, whose polarization is perhaps less problematic to determine (Kodaira and Yokoya, 2003). 
 
D) Lastly, the twist-2 T-odd functions $h_1^{\bot}$, corresponding to transverse polarization in an unpolarized nucleon, and the unpolarized distribution function $f^{\bot}_{1T}$ (Boer and Mulders, 1998) in a tranversely polarized nucleon find no place in parameterization (\ref{pargam}).

\subsection{Consequences on $g_1$ and $g_2$}

Now we examine some consequences of our results on the DIS structure functions $g_1(x)$ and $g_2(x)$, whose properties have been studied by various authors (Anselmino {\it et al.}, 1995; Jaffe and Ji, 1991b; Bluemlein and Tkablaze, 1999). To this end, here, and only in this subsection, we re-introduce the flavor indices, dropped out in formula (\ref{ht00}), in order to recover the usual definitions of those functions. Moreover, we recall that 
\begin{equation}
g_i(x) = \sum_a e_a^2 [g_i^a(x)+{\bar g}_i^a(x)] ~~~~ 
(i = 1,2; ~~~ a = u, d, s),
\end{equation}
where $e_a$ is the fractional charge of the flavour $a$ and the barred quantities refer to antiquarks. On the other hand,
\begin{equation}
g_T = g_1(x) + g_2(x) = g_{T,e}(x) + g_{T,o}(x). \label{sr}
\end{equation} 
Here we have defined
\begin{equation}
g_{T,e(o)}(x) = \sum_a e_a^2 \int d^2p_{\bot} [{g'}^a_{T,e(o)}(x,{\bf p}_\bot^2) 
+{{\bar g}'}^a_{T,e(o)}(x,{\bf p}_\bot^2)].
\end{equation}
But eq. (\ref{ff3}) implies
\begin{equation}
g_{T,e}(x) = \sum_a e_a^2 \frac{m_a}{xM}[h_1^a(x)+{\bar h}_1^a(x)] 
+ O(M^2/Q^2). \label{sre}
\end{equation} 

As discussed in subsect. 4.2, $g_{T,e}$ is negligibly small for a nucleon. Therefore our result is in contrast with the Burkhardt-Cottigham (1970) (BC) sum rule, {\it i. e.},
\begin{equation}
\int_0^1 g_2(x)dx = 0. \label{bc}
\end{equation}
Indeed, integrating both sides of eq. (\ref{sr}) between 0 and 1, and assuming relation (\ref{bc}), implies
\begin{equation}
\int_0^1 g_1(x)dx \approx \int_0^1 g_{T,o}(x)dx.
\end{equation}
But this result is unacceptable, since a twist-2, T-even function like $g_1(x)$ has {\it a priori} no relation with $g_{T,o}$, which is twist-3 and T-odd. 

Furthermore eq. (\ref{bc}) implies, together with the operator product expansion (Anselmino {\it et al.}, 1995),
\begin{equation}
g_1(x) + g_2(x) = \int_x^1\frac{dy}{y} g_1(y)+ g_T^{(3)}, 
\label{ww}
\end{equation}
where $g_T^{(3)}$ is the twist-3 contribution to $g_T$ (Anselmino {\it et al.}, 1995), to be identified, according to our results, with $g_{T,o}$. Then eq. (\ref{sr}) would yield
\begin{equation}
\int_x^1\frac{dy}{y} g_1(y) = g_{T,e}(x)+ O(M^2/Q^2),
\label{ww1}
\end{equation}
which appears in contrast with data of $g_1(x)$ (Ashman {\it et al.}, 1988, 1989; Airapetian {\it et al.}, 1998), enforcing arguments against the BC rule (See Anselmino {\it et al.} (1995) and articles cited therein).
An experimental confirmation of the violation of the BC rule was found years ago in a precision measurement of $g_2(x)$ (Anthony {\it et al.}, 2003). 

Also the Efremov-Leader-Teryaev (ELT) sum rule, according to the version given by Anselmino {\it et al.} (1995), {\it i. e.},  
\begin{equation}
\int_0^1 dx x [g_1(x)+2g_2(x)] = 0,
\label{ww2}
\end{equation}
is in contrast with our result. Indeed, it gives rise, together with eqs. (\ref{sr}) and (\ref{sre}), to the approximate relation
\begin{equation}
\int_0^1 dx x g_1(x) \approx \int_0^1 dx 2x g_{T,o}(x),
\label{ww3}
\end{equation}
which, again, relates a T-even function to a T-odd one. However, it is worth noting that the ELT sum rule was successively reformulated (Efremov {\it et al.}, 1997) by suitably redefining $g_1$ and $g_2$.

\section{Fragmentation Correlator}

The fragmentation correlator (\ref{corrp}) can be made gauge invariant analogously to the distribution correlator, {\it i. e.}, for a quark,
\begin{equation}
\Delta_{ij}(p; P,S) = 2{\cal P}\int\frac{d^4x}{(2\pi)^4} e^{ipx} 
\langle 0|{\cal L}(x) \bar{\psi}_j(0)a(P,S)a^\dagger(P,S)\psi_i(x)|0\rangle, \label{corrq}
\end{equation}
where ${\cal L}(x)$ is given by eq. (\ref{link}). The object (\ref{corrq}) may be treated analogously to the distribution correlator, according to the previous sections. Indeed, also in this case, for an antiquark one has to change the 
four-momentum from $p$ to $-p$ and to put a minus sign in front of the correlator. Moreover one has to choose the path ${\cal I}_+$ for quark fragmentation from $e^+e^-$ annihilation, whereas the path ${\cal I}_-$ refers to fragmentation in SIDIS. The only important difference with the distribution correlator is that one has to take into account also the nonperturbative interactions among the final hadrons produced. However, as we shall see in a moment, this does not involve any change in the parameterization. 

We treat only the case of pions, adopting for T-odd terms an approximation analogous to the one discussed in subsection 5.1, valid for small transverse momenta of the final hadron with respect to the fragmenting quark. Under this 
condition, we have
\begin{eqnarray}
\Delta (p) &=& 2p^+\{{\bar\Delta}^{(f)} (p)\delta(p^2-m^2)+{\bar\Delta}^{(i)} (p)[\pi(p^2-m^2+i\epsilon)]^{-1}\}, \label{fc}
\\
{\bar\Delta}^{(f)} (p) &=& \frac{1}{2}(p\hspace{-0.45 em}/+m)D_{\pi}, \label{fc0}
\\
{\bar\Delta}^{(i)} (p) &=& \gamma_-\gamma_+[p\hspace{-0.45 em}/_{\bot} D_{\pi}^{\bot} + \gamma_5 p\hspace{-0.45 em}/_{\bot}
{\bar S}\hspace{-0.65 em}/_{\bot}H']. \label{fc1}
\end{eqnarray}
Here $D_{\pi}$ is the common fragmentation function of the pion; $D_{\pi}^{\bot}$, defined according to Mulders and Tangerman (1996), is the analog of $f^{\perp}$; last, $H'$ assumes the role of the Collins (1993) function, describing the asymmetry of a pion fragmented from a 
transversely polarized quark, the so-called Collins asymmetry (see also Leader, 2004). 

Final state interactions give rise to terms which decrease as inverse powers of $Q$, independent of the nature of the interactions themselves. As an example, we re-consider the interactions which produce the above mentioned Collins asymmetry from a different point of view. Analogously to the distribution functions illustrated in remark D at subsect. 4.3, such an asymmetry may be connected to the absorptive part of an amplitude of the type $\langle + | -\rangle$, where $\pm$ denotes the helicity of the fragmenting quark. This kind of amplitude - a typical helicity flip one - behaves as
\begin{equation}
\langle + | -\rangle = {\cal B} sin\theta,
\end{equation}
where ${\cal B}$ is a given function, weakly dependent on the quark momentum. Then, similarly to eq. (\ref{ampl}), we conclude that the effect of the final state interaction between the fragmenting quark and the fragmented hadron decreases like $Q^{-1}$. This confirms our previous result, but independent of the nature of the interaction.  

More generally, we examine the interactions that the fragmented hadron, say hadron $B$, undergoes with other final hadrons. These cause in the momentum ${\bf P}_B$ of $B$ a change ${\bf \Delta P}_B$ which depends weakly on $Q$, since the multiplicity of the hadrons produced in inclusive reactions increases only logarithmically with energy. Moreover, for sufficiently large $Q$ and not too small fractional momenta $z$ of $B$ with respect to the fragmenting quark, the ratio  
\begin{equation}
{\cal R} = \frac{|{\bf \Delta P}_B|}{|{\bf P}_B|}
\end{equation}
is quite small. Then, under such conditions, ${\cal R}$ decreases approximately like $Q^{-1}$. Our result agrees with the approach by Collins and Soper (1981), who do not include "soft" final state interaction in the leading term of  (almost) back-to-back fragmentation in $e^+e^-$ annihilation.   

\section{Asymmetries}

In this section we consider some important azimuthal and single spin asymmetries, which, as is well known, may be produced by coupling two 
chiral-even or two chiral-odd TMD distribution or fragmentation functions. More precisely, the terms of the hadronic tensor which give rise to asymmetries are written as convolutive products of two "soft" functions times a suitable weight function (Boer {\it et al.}, 2000; Di Salvo, 2007a) which changes from asymmetry to asymmetry. These last depend on some azimuthal angle $\phi$, relative to the final hadron (for SIDIS and $e^+e^-$ annihilation), or to the final muon pair (for DY). Some of these asymmetries arise from the first order correction of the hadronic tensor, while others belong to the second order one, whose complete parameterization is not considered in this paper. 

 {\bf A) Cahn effect}

This effect, pointed out for the first time by Cahn (1978), has been exhibited by Anselmino {\it et al.} (2007) examining some SIDIS data (Arneodo {\it et al.}, 1987; Ashman {\it et al.}, 1991; Adams {\it et al.}, 1993) (see also Anselmino {\it et al.}, 2006). We consider the asymmetry corresponding to the "product"  
\begin{equation}
 A_C \propto f^{\bot}\otimes D_{\pi} + f_1\otimes D^{\bot}_{\pi}. \label{ch}
\end{equation} 
This asymmetry is proportional to $cos \phi$ and decreases like $Q^{-1}$.
To the extent that $f^{\bot}$ and $D^{\bot}_{\pi}$ can be approximated by $f_1$ and $D_{\pi}$ respectively, one speaks 
properly of Cahn effect (Anselmino {\it et al.}, 2007): this amounts to neglecting quark-gluon interactions, see eq. (\ref{ff2}) for 
distribution functions, an analogous equation holding for unpolarized fragmentation functions. This approximation is acceptable for relatively large $Q$ and at small $x$, as shown by Anselmino {\it et al.} (2007). However, one has to observe that both $f^{\bot}$ and $D^{\bot}_{\pi}$ are "hybrid" functions and in general their T-odd contributions cannot be neglected. 

It is worth considering also the "product"
\begin{equation}
 A_{C2} \propto f^{\bot}\otimes D^{\bot}_{\pi}, \label{ch2}
\end{equation}
which generates a $cos 2\phi$ asymmetry decreasing like $Q^{-2}$, hardly distinguishable from another one, arising from the "product" of two chiral-odd functions, as we shall see in a moment. Under the approximation just discussed, we predict a sort of "second order" Cahn effect.   

 {\bf B) Qiu-Sterman effect}

An important transverse single spin asymmetry is the one predicted by Qiu and Sterman (1991, 1992, 1998) (QS) (see also Efremov and Teryaev, 1984, 1985; Boer {\it et al.}, 1998, 2003b).  This can be observed both in SIDIS and in DY, by integrating over the transverse momentum of the final hadron detected (SIDIS) or of the final pair (DY). This is described by the "products"
\begin{equation}
 A_{QS} \propto g'_T\otimes D_{\pi}  ~~~~~ {\mathrm (in ~~ SIDIS) ~~ and} ~~~~~~~ \propto g'_T \otimes{\bar f}_1   +  c. c. ~~~~~ ({\mathrm in ~~ DY}), \label{qst}
\end{equation}
the "bar" indicating the antiquark function and $c. c.$ "charge conjugated".
A similar effect could be observed in $e^+e^-$ annihilation, if one of the final hadrons observed is spinning. This asymmetry decreases like $Q^{-1}$. Moreover, since $g'_T$ is prevalently T-odd, while $f_1$, ${\bar f}_1$  and 
$D_{\pi}$ are T-even, the asymmetry is expected to assume an opposite sign in SIDIS and DY. 

 {\bf C) Sivers effect}  

The Sivers (1990, 1991) single transverse spin asymmetry is described by the "product"
\begin{equation}
 A_{SIV} \propto f_T\otimes D_{\pi} ~~~~~ {\mathrm (in ~~ SIDIS) ~~ and} ~~~~~
 \propto f_T \otimes {\bar f}_1   + c. c. ~~ ({\mathrm in ~~ DY}). \label{sv}
\end{equation}

This asymmetry was detected by HERMES (Airapetian {\it et al.}, 2005b; 
Diefenthaler, 2005) and COMPASS (Alexakhin {\it et al.}, 2005) experiments.
It is T-odd, since it consists of the "product" of a T-odd function ($f_T$) times a T-even function ($D_{\pi}$ or ${\bar f}_1$). Therefore the asymmetry is predicted to change sign (Collins, 2002; Collins {\it et al.}, 2006; Anselmino {\it et al.}, 2009), according as to whether it is observed in SIDIS or DY, similar to the QS effect. 

However the T-odd character of $f_T$ leads us to conclude that the Sivers asymmetry decreases like $Q^{-1}$, in disagreement with the current literature (Boer and Mulders, 1998; Efremov {\it et al.}, 2006b; Anselmino {\it et al.}, 2007).

Furthermore the third eq. (\ref{fg}), {\it i. e.}, $f_T \approx g'_{T,o}$, implies, together with eqs. (\ref{qst}) and (\ref{sv}), that the Sivers and QS asymmetries are related to each other, although the weight functions (Boer {\it et al.}, 2000; Di Salvo, 2007a) involved in the two "products" are different.
This analogy was already noticed by other authors (Boer {\it et al.}, 2003b; Ji {\it et al.}, 2006a,b,c; Koike {\it et al.}, 2008). 

{\bf D) Collins effect and Boer-Mulders effect}

In the framework of chiral-odd functions, an important single spin asymmetry is produced by combination of two transversities. In particular, single transverse polarization gives rise to an asymmetry described by the "product"
\begin{eqnarray}
 A_{COL} &\propto& h_{1T}\otimes H'  ~~~~~ {\mathrm (in ~~ SIDIS), ~~ or}  \label{col} \\
 ~~~~~~~ A_{BM} &\propto& h_{1T} \otimes{\bar h}' + c. c. ~~~ ({\mathrm in ~~ DY}). \label{cm}
\end{eqnarray}
The asymmetry $ A_{COL}$ - predicted by Collins (1993) and exhibited by HERMES data (Airapetian {\it et al.}, 2005b; Diefenthaler, 2005) - decreases like $Q^{-1}$ according to our treatment. It has been studied recently by Leader (2004), Anselmino (2009, 2010) and Boer (2009).

We have also the following azimuthal, $cos 2\phi$ asymmetries: 
\begin{eqnarray}
 A_{CL2} &\propto& h'\otimes H'  ~~~~ {\mathrm (in ~~ SIDIS), ~~ or} \\
 A_{BM2} &\propto& h' \otimes{\bar h}' ~~~~ {\mathrm (in ~~ DY), ~~ or} \\
A_{CL3}  &\propto& H'\otimes {\bar H}'  ~~~~ {\mathrm (in ~~ e^+e^- ~~~ annihilation)},\label{coa}
\end{eqnarray}
which decrease like $Q^{-2}$. Therefore, as in the case of the Sivers asymmetry, we obtain a $Q^2$ dependence of asymmetries (\ref{col}) to (\ref{coa}) which differs from other authors (Boer and Mulders, 1998; Efremov {\it et al.}, 2006a; Burkardt and Hannafious, 2008). Our prediction for the Boer-Mulders asymmetry $A_{BM2}$ is supported (Di Salvo, 2007a) by DY data (Falciano {\it et al.}, 1986; Guanziroli {\it et al.}, 1988; Conway {\it et al.}, 1989). On the other hand, the $Q^2$ dependence of the Collins and Sivers asymmetries might be tested in new planned experiments at higher energies (Afanasev {\it et al.}, Jefferson Lab., hep-ph/0703288, 2007).
  
\section{Summary}
				
 In the present paper we have studied the gauge invariant quark-quark correlator, which we have expanded in powers of the coupling and split into a T-even and a T-odd part. Working in the KS gauge, the Politzer theorem on EOM has 
allowed us to interpret each term of the expansion according to Feynman-Cutkosky graphs, involving higher correlators and corresponding to powers of $gM/Q$. We have also elaborated an algorithm for writing a gauge invariant sector of the hadronic tensor in deep inelastic processes, like SIDIS, DY and $e^+e^-$ annihilation. This gives rise to a rather long and complicated sum of terms. However, in the gauge considered, and especially at small transverse momenta, the "Born" terms of the type (\ref{ht00}) prevail over the remaining ones, as we have shown explicitly for first order correction in $gM/Q$.

The zero order term and the first order correction of the expansion have been examined in detail. In both cases the Politzer theorem produces a considerable reduction of independent functions with respect to the naive parameterization 
in terms of Dirac components, giving rise to approximate (up to powers of $gM/Q$) relations among "soft" functions. These relations survive QCD evolution. One such relation has been approximately verified against experimental 
data (Airapetian {\it et al.}, 2005b; Avakian {\it et al.}, 2005), another one suggests a method for determining approximately transversity, while others could be checked in next experiments (Bunce {\it et al.}, 2000; Adams {\it et al.}, 1993). Also an energy scale, introduced in the naive parameterization for dimensional reasons, has been determined in our approach, leading to predictions on $Q^2$ dependence of various azimuthal asymmetries. One of these predictions finds confirmation in unpolarized DY data (Falciano {\it et al.}, 1986; Guanziroli {\it et al.}, 1988; Conway {\it et al.}, 1989). The hierarchy 
of TMD functions in terms of inverse powers of $Q$ is established taking into account not only the Dirac operators, as in the case of common functions (Jaffe and Ji, 1991a, 1992), but also the $p_{\bot}$ dependence, since in this case the orbital angular momentum plays a role as well as spin.  

Moreover a relation is found among $g_T$, the QS asymmetry and the Sivers asymmetry; in particular, both $g_T$ and the two asymmetries are found to change sign according as to whether they are observed in SIDIS or in DY. We draw also some conclusions about the structure function $g_2(x)$, in particular against the BC sum rule. 

Quark fragmentation involves "soft" interactions among final hadrons, but this does not imply a substantial difference with the distribution correlator. Rather, a caveat should be kept in mind for timelike photons, in DY and $e^+e^-$ 
annihilation, when $Q$ approaches the energy of a vector boson resonance, like the $\Upsilon$ or the $Z^0$. Since such a resonance interferes with the photon, one has to take into account its offshellness, quite different	than $Q^2$. A particular attention has to be paid also to the case when the active quark (or antiquark) comes from gluon annihilation, as occurs, for example, in DY from proton-proton collisions. This may give rise to T-odd Feynman-Cutkosky 
graphs, in which the (anti-)quark propagator is only slightly off-shell. These two situations deserve a separate treatment.

As a conclusion, we stress that, although other authors, like EFP, LT and Efremov and Teryaev (1984) already proposed, years ago, a decomposition of the hadronic tensor in terms of Feynman-Cutkosky graphs, our deduction, based on EOM, leads to strong constraints on the parameterization of the "soft" parts of the graphs.    
				  
\vskip 0.30in

\section*{Acknowledgments}
The author is grateful to his friends A. Blasi, A. Di Giacomo and N. Paver for fruitful discussions.
\vskip 0.30in
\vspace {10pt}

\setcounter{equation}{0}
\renewcommand\theequation{A. \arabic{equation}}

\appendix{\large \bf Appendix A}

We deduce a recursion formula for the terms of the expansion of the correlator. Our starting point is the Politzer (1980) theorem, which implies
\begin{equation}
\langle P,S|{\bar \psi}_j(0){\cal L}(x) (iD\hspace{-0.7em}/-m)_{il} \psi_l(x)|P,S\rangle = 0. \label{pol1}
\end{equation}
Here $|P,S\rangle$ denotes the state of a hadron (for instance, but not necessarily, a nucleon) with four-momentum $P$ 
and PL four-vector $S$. $\psi$ is the quark field, of which we omit the color and flavor index. $D_{\mu} = \partial_{\mu} - ig {\bf A}_{\mu}$ is the covariant derivative, adopting for the gluon field the shorthand notation ${\bf A}_{\mu}$ for $A^a_{\mu}\lambda_a$. For the sake of simplicity, color and flavor indices of the quark field have been omitted.
Moreover 
\begin{equation}
{\cal L}(x) = \sum_{n=0}^{\infty} (ig)^n\Lambda_n(x), \label{link1}
\end{equation} 
where $g$ is the strong coupling, while $\Lambda_0(x) = 1$. We have, for $n\geq 1$, in the KS gauge, 
\begin{equation}
\Lambda_n(x) = \int_{x_{1}}^{x_{2}}dz_1^{\mu_1} \int^{z_{1}}_{x_{1}} dz_2^{\mu_2} ... 
\int^{z_{n-1}}_{x_{1}}dz_n^{\mu_n} \left[{\bf A}_{\mu_1}(z_1)
{\bf A}_{\mu_2}(z_2) ... {\bf A}_{\mu_n}(z_n)\right].\label{cor1a}
\end{equation}
Here we have adopted the reference frame and the notations and definitions introduced in sect. 2. In particular, $x_2$ is related to $x$: $x_2 \equiv 
(\pm\infty, x^+, {\bf x_{\bot}})$, $x \equiv (x^-, x^+, {\bf x_{\bot}})$.
It is worth observing that
\begin{equation}
\partial_{\mu}\Lambda_n = {\bf A}_{\mu}(x_{2})\Lambda_{n-1}. \label{deriv}
\end{equation}
Substituting expansion (\ref{link1}) into eq. (\ref{pol1}), we get
\begin{equation}
\sum_{n=0}^{\infty} (ig)^n  \left\{{\bar \psi}_j(0)\Lambda_n(x)
(i\partial\hspace{-0.55em}/-m)_{il} \psi_l(x)
-i{\bar \psi}_j(0)\Lambda_{n-1}(x) [i{\bf A}\hspace{-0.6 em}/ (x)]_{il}\psi_l(x)\right\} = 0, \label{ser}
\end{equation}
with
\begin{equation}
\Lambda_{-1}(x) = 0 ~~~~ {\mathrm and} ~~~~~ \Lambda_0(x) = 1. \label{teq}
\end{equation}
Eq. (\ref{ser}) is an operator equation, to be intended in a weak sense: it holds when calculated between hadronic 
states. All equations of this Appendix will be of this type from now on. 

Looking for a perturbative solution for the correlator in powers of $g$, we set each term of the series (\ref{ser}) equal to zero, {\it i. e.},
\begin{equation}
(i\partial\hspace{-0.55em}/-m){\cal O}_n(x) = i{\bf A}\hspace{-0.6em}/(x)
{\cal O}_{n-1}(x), \label{opr}
\end{equation}
where
\begin{equation}
[{\cal O}_n(x)]_{ij} = {\bar \psi}_j(0)\Lambda_n(x)\psi_i(x).
\end{equation}
By Fourier transforming both sides of eq. (\ref{opr}), and recalling relation (\ref{deriv}), we get 
\begin{equation}
(p\hspace{-0.45 em}/-m){\tilde{\cal O}}_n(p) = i \gamma_{\mu} \int\frac{d^4x}{2\pi^4}e^{ipx} \left[{\bf A}^{\mu}(x_2){\cal O}_{n-1}(x) + {\cal O}_{n-1}(x) {\bf A}^{\mu}(x)\right], \label{rr}
\end{equation}
where 
\begin{equation} 
{\tilde{\cal O}}_n(p) = \int\frac{d^4x}{2\pi^4}e^{ipx} {\cal O}_n(x).
\end{equation}
Eq. (\ref{rr}) can be rewritten as
\begin{equation} 
(p\hspace{-0.45 em}/-m){\tilde{\cal O}}_n(p) = i\gamma_{\mu}\int\frac{d^4k}{2\pi^4}
\left[{\tilde{\bf A}}^{\mu}(k){\tilde{\cal O}}_{n-1}(p-k) + {\tilde{\cal O}}_{n-1}(p-k) {\hat{\bf A}}^{\mu}(k)\right], 
\label{recur}
\end{equation}
where
\begin{eqnarray}
{\hat{\bf A}}^{\mu}(k) &=& \int\frac{d^4x}{2\pi^4}e^{ikx} {\bf A}^{\mu}(x), 
\\
{\tilde{\bf A}}^{\mu}(k) &=& \delta(k^+) \displaystyle\lim_{M\to\infty}\int d\kappa e^{-i\kappa M}{\hat {\bf 
A}}_{\mu}(k^-,\kappa, {\bf k}_{\perp}). 
\end{eqnarray}
Eq. (\ref{recur}) is a recursion formula for ${\tilde{\cal O}}_n(p)$, eqs. (\ref{teq}) constituting the first steps. This formula implies eqs. (\ref{hom0}) (for $n$ = 0) and (\ref{pol}) (for $n\geq 1$) in the text. In particular, as 
regards eq. (\ref{pol}), the quantity $\Gamma_n$ results in
\begin{equation}
\Gamma_n =  N\langle P,S|{\tilde{\cal O}}_n(p)|P,S\rangle, \label{gn}
\end{equation}
where $N$ is a normalization constant. The operator ${\tilde{\cal O}}_n(p)$ in eq. (\ref{recur}) corresponds to a graph endowed with $n$ gluons, such that the $n$-th gluon leg is attached to the quark leg on the left side of the graph (see figs. 2a and 3a).

Taking into account the hermitian character of ${\hat{\bf A}}^{\mu}(k)$ and the relation $[{\tilde{\cal O}}_n(p)]^\dagger = \gamma_0{\tilde{\cal O}}_n(p)\gamma_0$, eq. (\ref{recur}) implies 
\begin{equation} 
{\tilde{\cal O}}_n(p)(p\hspace{-0.45 em}/-m) = -i\int\frac{d^4k}{2\pi^4}
[{\tilde{\cal O}}_{n-1}(p-k){\tilde{\bf A}}^{\mu}(k) + {\hat{\bf A}}^{\mu}(k) {\tilde{\cal O}}_{n-1}(p-k)]\gamma_{\mu}. \label{recur1}
\end{equation}
In this case ${\tilde{\cal O}}_n(p)$ corresponds again to a graph with $n$ gluons, but such that the $n$-th gluon is attached to the quark leg on the right side of the graph. This last result implies that $\Gamma_n$ represents any graph 
with $n$ gluons, each gluon leg being attached to the left or right quark leg.

\vskip 0.30in

\setcounter{equation}{0}
\renewcommand\theequation{B. \arabic{equation}}

\appendix{\large \bf Appendix B}
 
Here we deduce the parameterizations of the quark-quark correlator at zero order and of the quark-gluon-quark correlation, arising from first order correction.

{\bf B.1. ~~~ The Zero Order Quark-Quark Correlator} 

The matrix $\Gamma_0(p)$, defined by
\begin{equation}
(\Gamma_0)_{ij} = N\int\frac{d^4x}{(2\pi)^4} e^{ipx} 
\langle P,S|\bar{\psi}_j(0) \psi_i(x)|P,S\rangle,  \label{co0a}
\end{equation}
fulfils the homogeneous Dirac equation
\begin{equation}
(p\hspace{-0.45em}/-m)\Gamma_0(p) = 0, \label{dirac}
\end{equation}
where $m$ is the rest mass of the quark. As shown in Appendix A, this is a consequence of the Politzer theorem. This implies, at zero order in the coupling,
\begin{equation}
(\partial\hspace{-0.55em}/-m)\psi(x) = 0. \label{dirac0}
\end{equation}
Therefore, in the approximation considered, the quark can be treated as if it were on shell (see also Qiu, 1990). Then, initially, we consider the Fourier expansion of the unrenormalized field of an on-shell quark, {\it i. e.},
\begin{equation} 
\psi(x) = \int \frac{d^3{\tilde p}}{(2\pi)^{3/2}} \frac{1}{\sqrt{2{\cal P}}} e^{-ipx} \sum_s u_s(p) c_s(p). \label{field}
\end{equation} 
Here $s = \pm 1/2$ is the spin component of the quark along a given direction in the quark rest frame, $u$ its four-spinor, $c$ the destruction operator for the flavor considered and
\begin{equation}
d^3{\tilde p} = d^4p ~ \delta\left(p^--\frac{m^2+{\bf p}^2_{\perp}}{2p^+}\right), ~~~~~ \ ~~~~~ \ ~~~~~ \ ~~~~ {\cal P} 
= p^+/\sqrt{2}. \label{eqn}
\end{equation}
As regards the normalization of $u_s$ and $c_s$, we assume
\begin{equation} 
{\bar u}_s u_s = 2m, ~~~~~~~~ \ \langle P,S|c_s^{\dagger}(p')c_s(p)|P,S\rangle = (2\pi)^3\delta^3({\tilde {\bf p}'}-{\tilde {\bf p}}) q_s(p), \label{norm}
\end{equation}
where
\begin{equation} 
{\tilde {\bf p}}\equiv(p^+,{\bf p}_{\perp})\label{def2}
\end{equation}
and $q_s(p)$ is the probability density to find a quark with spin component $s$ and four-momentum $p \equiv (p^-, {\tilde {\bf p}})$, with $p^- = (m^2+{\bf p}^2_{\perp})/{2p^+}$. For an antiquark the definition is analogous, except 
that, in the Fourier expansion (\ref{field}), we have to substitute the destruction operators $c_s$ with the creation 
operators $d^{\dagger}_s$ and $p$ with $-p$ in the exponential.
 
Choosing the quantization axis along the hadron momentum ${\bf P}$ in the frame
defined at the beginning of sect. 4, and substituting eq. (\ref{field}) into eq. (\ref{co0a}), we get
\begin{eqnarray}
&~&(\Gamma_0)_{ij}(p) =\frac{N}{2{\cal P}} \sum_{s,s'}\int\frac{d^3{\tilde p'}}{(2\pi)^3} \langle P,S|c_s^{\dagger}(p) 
c_{s'}(p')|P,S\rangle \nonumber
\\
&\times&[u_{s'}(p')]_i [\bar{u}_{s}(p)]_j ~ \delta\left(p^--\frac{m^2+{\bf p}^2_{\perp}}{2p^+}\right). \label{dm1}
\end{eqnarray}
But owing to the second eq. (\ref{norm}) we have
\begin{equation}
\Gamma_0(p) = [\Gamma_0^a(p) + \Gamma_0^b(p)] ~ \delta\left(p^--\frac{m^2+{\bf p}^2_{\perp}}{2p^+}\right), \label{dm2}
\end{equation}
where
\begin{eqnarray}
\Gamma_0^a(p) &=& \frac{N}{2{\cal P}}\sum_{s}  
\langle P,S|c_{s}^{\dagger}(p) c_{s}(p)|P,S\rangle u_{s}(p) \bar{u}_{s}(p), \label{dm2a}
\\
\Gamma_0^b(p) &=& \frac{N}{2{\cal P}}\sum_{s}  
\langle P,S|c_{-s}^{\dagger}(p) c_{s}(p)|P,S\rangle u_{-s}(p) \bar{u}_{s}(p). \label{dm2b}
\end{eqnarray}
Firstly we elaborate $\Gamma_0^a$. We have
\begin{equation}
u_{s}(p)\bar{u}_{s}(p) =  \frac{1}{2}(p\hspace{-0.45 em}/+m)(1+2s\gamma_5S\hspace{-0.65 em}/^a_{\parallel}). 
\label{dd22}
\end{equation}
Here $S^a_{\parallel}$ is a four-vector such that, in the quark rest frame, 
$S^a_{\parallel}\equiv(0,\lambda/|\lambda|{\hat{\bf P}})$, $\lambda = {\bf S}\cdot {\hat{\bf P}}$, ${\hat{\bf P}} = 
{\bf P}/|{\bf P}|$ and ${\bf S}$ is the unit spin vector of the hadron in its rest frame. Therefore
\begin{equation}
\Gamma_0^a(p) =
\frac{N}{2{\cal P}}\frac{1}{2}(p\hspace{-0.45 em}/+m)\left[f_1(p)+ \Delta' q(p)\gamma_5 S\hspace{-0.65 
em}/^a_{\parallel}\right],\label{long}
\end{equation}
where
\begin{equation}
f_1(p) = \sum_{s}\langle P,S|c_{s}^{\dagger}(p) c_{s}(p)|P,S\rangle
\end{equation}
is the unpolarized transverse momentum distribution of the quark, while 
\begin{equation}
\Delta'q(p) = \sum_{s}2s\langle P,S|c_{s}^{\dagger}(p) c_s(p)|P,S\rangle. \label{lpol}
\end{equation}
According to transformation properties of one-particle states under rotations, one has
\begin{equation}
|P,S\rangle = cos\frac{\theta}{2}|P,+\rangle + i |P,-\rangle sin\frac{\theta}{2}, \label{rot}
\end{equation}
where $\pm$ denotes the (positive or negative) helicity of the hadron and $\theta$ the angle between ${\bf P}$ and 
${\bf S}$. Substituting eq. (\ref{rot}) into eq. (\ref{lpol}), and taking into account parity conservation, we get
\begin{equation}
\Delta'q(p) = cos\theta g_{1L}(p). \label{lpol0}
\end{equation}
Here
\begin{equation}
g_{1L}(p) = \sum_{s}2s\langle P,+|c_{s}^{\dagger}(p) c_{s}(p)|P,+\rangle = -\sum_{s}2s\langle P,-|c_{s}^{\dagger}(p) 
c_{s}(p)|P,-\rangle. \label{lpol1}
\end{equation}
is the longitudinally polarized TMD distribution of the quark, the last equality following from parity conservation.

Now we consider $\Gamma_0^b$. Eq. (\ref{rot}) yields, for $\theta = \pi/2$,
\begin{equation}
|\uparrow(\downarrow)\rangle = \frac{1}{\sqrt{2}}(|+\rangle \pm i|-\rangle), \label{rotq}
\end{equation}
where $|\pm\rangle$ and $|\uparrow(\downarrow)\rangle$ denote quark states with spin components, respectively, along ${\hat{\bf P}}$ and along 
\begin{equation}
{\bf S}_{\perp}={\bf S}-\lambda{\hat{\bf P}}.
\end{equation}
Substituting eqs. (\ref{rot}) and (\ref{rotq}) into eq. (\ref{dm2b}), and taking into account again parity conservation, we get
\begin{equation}
\Gamma_0^b(p) = \frac{N}{2{\cal P}}\frac{1}{2}sin\theta  h_{1T}(p)
(|\uparrow\rangle \langle\uparrow| - |\downarrow\rangle \langle\downarrow|), \label{tr0}
\end{equation}
where
\begin{equation}
h_{1T}(p) = \langle P,-|c_+^{\dagger}(p) c_-(p)|P,+\rangle = \langle P,+|c_-^{\dagger}(p) c_+(p)|P,-\rangle
\end{equation}
is the TMD transversity of the quark. Returning to the Dirac notation, we have 
\begin{equation}
|\uparrow\rangle \langle\uparrow| = \frac{1}{2}(p\hspace{-0.45 em}/+m)\left(1 + \gamma_5 S\hspace{-0.65 
em}/^b_{\perp}\right), ~~~~~~~~~ ~~~~~~~~~ |\downarrow\rangle \langle\downarrow| = \frac{1}{2}(p\hspace{-0.45 
em}/+m)\left(1 - \gamma_5 S\hspace{-0.65 em}/^b_{\perp}\right),
\end{equation}
where $S^b_{\perp}$ is such that $S^b_{\perp} \equiv (0,{\hat{\bf n}})$ in the quark rest frame and 
\begin{equation}
{\hat{\bf n}} = \frac{{\bf S}_{\perp}}{|{\bf S}_{\perp}|}.
\end{equation}
 Then eq. (\ref{tr0}) goes over into
\begin{equation}
\Gamma_0^b(p; P,S) = \frac{N}{2{\cal P}}\frac{1}{2}sin\theta \Delta_T q(p)
(p\hspace{-0.45 em}/+m)\gamma_5 S\hspace{-0.65 em}/^b_{\perp}. \label{tr1}
\end{equation}
Substituting eqs. (\ref{long}), (\ref{lpol0}) and (\ref{tr1}) into eq. (\ref{dm2}) yields
\begin{equation}
\Gamma_0 = \frac{N}{2{\cal P}}\frac{1}{2}(p\hspace{-0.45 em}/+m)\left[f_1+ g_{1L}\gamma_5 S\hspace{-0.65 em}/^q_{\parallel}+h_{1T} \gamma_5 S\hspace{-0.65 em}/^q_{\perp}\right]\delta\left(p^--\frac{m^2+{\bf p}^2_{\perp}}{2p^+}\right), \label{cc}
\end{equation}  
having set $S^q_{\parallel} = S^a_{\parallel}cos\theta$ and $S^q_{\perp} = S^b_{\perp}sin\theta$. Eq. (\ref{cc}) is a solution to eq. (\ref{dirac}), which is a consequence of the Politzer theorem at zero order in $g$. Since this equation survives renormalization - which generally implies only a weak $Q$-dependence (Sterman, 2005; Dokshitzer {\it et al.}, 1980) - the structure of $\Gamma_0$ is not changed by QCD evolution.

Lastly we deduce the expressions of the four-vectors $S^q_{\parallel}$ and $S^q_{\perp}$ in the frame where the quark momentum is ${\bf p}$. In the quark rest frame we have
\begin{equation}
S^q_{\parallel} \equiv (0, \lambda{\hat{\bf P}}), ~~~~~~~~~~~~ S^q_{\bot} \equiv (0, {\bf S}_{\bot}). \label{s0}
\end{equation} 
In view of the Lorentz boost, it is convenient to further decompose $\lambda{\hat{\bf P}}$ and ${\bf S}_{\bot}$ into 
components parallel and perpendicular to the quark momentum. We have
\begin{eqnarray}
\lambda{\hat{\bf P}} &=& \lambda cos\alpha{\hat{\bf p}} + {\bf\Sigma}_{\|}, ~~~~~~~~~ 
{\bf\Sigma}_{\|} = -cos\alpha\frac{{\bf p}_{\bot}}{|{\bf p}|}+ sin^2\alpha {\hat{\bf P}},
\\
{\bf S}_{\perp} &=& \lambda_{\bot}{\hat{\bf p}}+{\bf\Sigma}_{\bot}, ~~~~~~~~~ {\bf\Sigma}_{\bot} = |{\bf S}_{\perp}|cos\beta(cos\beta{\hat{\bf n}}-sin\beta{\hat{\bf k}}),
\end{eqnarray}
where 
\begin{eqnarray}
{\hat{\bf p}} &=& \frac{{\bf p}}{|{\bf p}|},  ~~~~~ \ ~~~~~~~ \ ~~~~~ \ ~~~~~~~ {\hat{\bf k}} = {\hat{\bf n}}\times\frac{{\hat{\bf p}}\times{\hat{\bf n}}}
{|{\hat{\bf p}}\times{\hat{\bf n}}|},
\\
\alpha &=& arccos ({\hat{\bf P}}\cdot{\hat{\bf p}}) ~~~~~ {\mathrm and} ~~~~~ \beta = arcsin ({\hat{\bf n}}\cdot{\hat{\bf p}}).
\end{eqnarray} 
The boost which transforms the four-momentum of the quark from $(m,0)$ to $(E,{\bf p})$, with $E = \sqrt{m^2+{\bf 
p}^2}$, changes only the components along ${\hat{\bf p}}$ of $\lambda{\hat{\bf P}}$ and of ${\bf S}_{\perp}$. In particular, the boost transforms the four-vector $(0,{\bf p})$ to ${\bar p}/m$, with ${\bar p}\equiv(|{\bf p}|, E{\hat{\bf p}})$. Therefore, since $\alpha$ and $\beta$ are $O(|{\bf p}_{\perp}|/|{\bf p}|)$ and $|{\bf p}|/{\cal P} = 
O(1)$, eqs. (\ref{s0}) go over into 
\begin{equation}
S^q_{\parallel} = \lambda \left(\frac{\bar{p}}{m}-\bar{\eta}_{\perp}\right) + O(\bar{\eta}^2_{\perp}), ~~~~~~~~~~~ \ 
~~~~~~ S^q_{\perp} =  S_{\perp}+\bar{\lambda}_{\perp}\frac{\bar{p}}{m_q}+ O(\bar{\eta}^2_{\perp}),  \label{qspin1}
\end{equation}
where $\bar{\eta}_{\perp}=p_{\bot}/{\cal P}$ and $\bar{\lambda}_{\perp}= -S\cdot\bar{\eta}_{\perp}$.

{\bf B.2. ~~~ The Quark-Gluon-Quark Correlator}

Now we deduce a parameterization for the quark-gluon-quark correlator, defined by
\begin{equation}
\left[\Phi^{(1)}_{\mu}(p, k)\right]_{ij} = N\int\frac{d^4x}{(2\pi)^4} e^{i(p-k)x} \langle P,S|\bar{\psi}_j(0)[{\hat 
{\bf A}}_{\mu}(k)+{\tilde {\bf A}}_{\mu}(k)]\psi_i(x)|P,S\rangle.   \label{rec1}
\end{equation}
As shown in Appendix A, the Politzer theorem implies, at order 1 in the coupling, 
\begin{equation}
(p\hspace{-0.45 em}/-k\hspace{-0.47 em}/-m)\Phi^{(1)}_{\mu}(p,k) = 0, \label{eqd}
\end{equation}
which holds also after renormalization. Therefore our line of reasoning is the same as for $\Gamma_0$, that is, we start from unrenormalized fields and we take on-shell quarks, whose field satisfies expansion (\ref{field}). 
Substituting this expansion into eq. (\ref{rec1}), we get
\begin{eqnarray}
\Phi^{(1)}_{\mu}(p, k) &=& \Psi_{\mu}(p, k)\delta\left(p_1^--\frac{{m}^2+{\bf p}^2_{1\perp}}{2p_1^+}\right), ~~~~~~~~~ 
\label{fi1}\\
\Psi_{\mu}(p, k) &=& N\int\frac{d^3{\tilde p}'}{(2\pi)^3}\frac{1}{2\sqrt{{\cal P}_1{\cal P}'}} \sum_{s,s'}{\cal 
A}_{s,s',\mu}(p', k)u_s(p_1){\bar u}_{s'}(p').   \label{rec2}
\end{eqnarray}
Here $d^3{\tilde p}'$ and ${\cal P}'$ are defined analogously to eqs. (\ref{eqn}),
\begin{equation}
p_1 = p-k,  ~~~~~~ \ ~~~~~~~ \ ~~~~~ {\cal P}_1 = p_1^+/\sqrt{2}
\end{equation}
and
\begin{equation}
{\cal A}_{s,s',\mu}(p',k) = \langle P,S|c_{s}^\dagger(p')[{\hat {\bf A}}_{\mu}(k)+{\tilde {\bf A}}_{\mu}(k)] 
c_{s'}(p_1)|P,S\rangle. \label{c5}
\end{equation} 
Moreover the matrix element (\ref{c5}) fulfils a relation of the type 
\begin{equation} 
{\cal A}_{s,s',\mu}(p', k) = (2\pi)^3{\cal C}_{s,s',\mu}(p', k)\delta^3({\tilde {\bf p}}'-{\tilde {\bf p}}_1-{\tilde 
{\bf k}}),
\end{equation}
where ${\cal C}_{s,s',\mu}(p', k)$ is a quark-gluon correlator and ${\tilde {\bf p}}'$, ${\tilde {\bf p}}_1$ and 
${\tilde {\bf k}}$ are defined by eq. (\ref{def2}). Then eq. (\ref{rec2}) yields
\begin{equation}
\Psi_{\mu}(p, k) = \frac{N}{2\sqrt{{\cal P}_1{\cal P}}} \sum_{s,s'}{\cal C}_{s,s',\mu}(p,k)u_s(p_1){\bar u}_{s'}(p_0) \label{rec3a}
\end{equation}
and 
\begin{equation}
p_0\equiv(p_0^-,{\tilde {\bf p}}), ~~~~~ \ ~~~~~ \ ~~~~ p_0^- = \frac{{\bf p}^2_{\bot}+m^2}{2p^+}. \label{4mom}
\end{equation}
We rewrite eq. (\ref{rec3a}) as 
\begin{equation}
\Psi_{\mu}(p, k) = \frac{N}{2\sqrt{{\cal P}_1{\cal P}}} (\Psi_{\mu}^a + \Psi_{\mu}^b), \label{rec3b}
\end{equation}
where 
\begin{eqnarray}
\Psi_{\mu}^a &=& \sum_s{\cal C}_{s,s,\mu}(p_1, k)u_s(p_1){\bar u}_s(p_0),
\\
\Psi_{\mu}^b &=& \sum_s{\cal C}_{s,-s,\mu}(p_1, k)u_s(p_1){\bar u}_{-s}(p_0).
\end{eqnarray}
Taking into account the appropriate Lorentz transformations for the spinors involved, we have  
\begin{eqnarray}
u_s(p_1){\bar u}_s(p_0) &=& \frac{1}{2}(p\hspace{-0.45 em}/_1+m) U(p_1,p_0)(1+2s\gamma_5S\hspace{-0.65 em}/^q_{0\|})\label{rec4}, 
\\
u_s(p_1){\bar u}_{-s}(p_0) &=& \frac{1}{2}(p\hspace{-0.45 em}/_1+m) U(p_1,p_0) \gamma_5 (cos\chi S\hspace{-0.65 
em}/^q_{0\bot}+sin\chi {\bar S}\hspace{-0.65 em}/_{\bot}).\label{rec4a}  
\end{eqnarray}
Here
\begin{equation}
U(p_1,p_0) = exp\left[\frac{1}{2}(\phi_1{\hat {\bf p}}_1-\phi_0{\hat {\bf p}}_0)\cdot{\vec \alpha}\right],  ~~~~~~~~ \ ~~~~~~ \ ~~~~~ \ ~~~~~~~
\label{rec4b}  
\end{equation}
\begin{eqnarray}
\phi_1 &=& ln \frac{E_1+|{\bf p}_1|}{m}, ~~~~~~~~ \ ~~~~~~ \ ~~~~~ \ ~~~~~~ \ {\hat {\bf p}}_1 = \frac{{\bf p}_1}{|{\bf p}_1|},
\\
{\bf p}_1 &\equiv& ({\bf p}_{1\bot}, \frac{1}{\sqrt{2}}(p_1^+-p_1^-)), ~~~~~ \ ~~~~~~~~ \ ~~~~~ \ ~~~~~~ \ E_1 = \sqrt{{\bf p}_1^2+m^2},
\end{eqnarray}
analogous definitions holding for $\phi_0$ and ${\hat {\bf p}}_0$.
Moreover $S^q_{0\|}$ and $S^q_{0\bot}$ refer to the PL vector of a quark with four-momentum $p_0$, directly connected with nucleon polarization; they can be related to the nucleon longitudinal and transverse PL vectors, using the formulae elaborated at the end of sect. B1. ${\bar S}_{\bot}$ refers to the spin caused by spin-orbit coupling,
\begin{equation}
\sqrt{|p_{0\bot}^2|}{\bar S}_{\bot\alpha} = \epsilon_{\alpha\beta\gamma\rho} n_+^{\beta}n_-^{\gamma} p_{0\bot}^{\rho}.
\end{equation}
Last, $\chi$ is a real, "soft" parameter, which in general will depend on $p_0$ and $p_1$; it will be included in the definitions of two of the "soft" 
correlation functions.

We assume $\theta_0, \theta_1 <<1$, where $\theta_0$ and $\theta_1$ are, respectively, the angle between ${\bf p}_0$ 
and ${\bf P}$ and the one between ${\bf p}_1$ and ${\bf P}$. Then
\begin{equation}
U(p_1,p_0) \approx cosh \varphi + \frac{1}{2\varphi}
\gamma_0(\gamma_3a +\gamma_i r_{\bot}^i)sinh \varphi, \label{rec4c}  
\end{equation}
with
\begin{eqnarray}
\varphi &=& \frac{1}{2}\sqrt{(\phi_0-\phi_1)^2+\theta^2\phi_0\phi_1}, ~~~~~~ \ ~~~~~~ \ ~~~~~~ \ ~~~~
\theta = \theta_1-\theta_0,\\
a &=& \phi_1-\phi_0- \frac{1}{2}(\phi_1\theta_1^2-\phi_0\theta_0^2), ~~~~~~ \ ~~~~~~ 
{\bf r_{\bot}} = \frac{\phi_1}{|{\bf p}_1|}{\bf p}_{1\bot} - \frac{\phi_0}{|{\bf p}_0|}{\bf p}_{0\bot,} 
\label{rec4d}  
\end{eqnarray}

Then $\Psi_{\mu}$ results in
\begin{equation}
\Psi_{\mu}(p_1, k) \approx \frac{1}{2}(p\hspace{-0.45 em}/_1+m) {\cal L} 
[{\cal C}_{\mu}+\Delta{\cal C}_{\mu}\gamma_5 S\hspace{-0.65 em}/^q_{\|}+
\Delta_T{\cal C}_{\mu}\gamma_5 S\hspace{-0.65 em}/^q_{\bot}+
\Delta_T{\cal C}'_{\mu}\gamma_5 {\bar S\hspace{-0.65 em}/}_{\bot}], 
\label{rec5}  
\end{equation}
where
\begin{equation}
{\cal L} = \frac{N}{2\sqrt{{\cal P}_1{\cal P}}}[cosh \varphi + \frac{1}{2\varphi} \gamma_0(\gamma_3a +\gamma_i r_{\bot}^i)sinh \varphi]
\end{equation}
and
\begin{eqnarray}
{\cal C}_{\mu}(p_1, k) &=& \sum_s {\cal C}_{s,s,\mu}(p_1, k),
\\
\Delta{\cal C}_{\mu}(p_1, k) &=& \sum_s 2s {\cal C}_{s,s,\mu}(p_1, k),
\\
{\Delta_T\cal C}_{\mu}(p_1, k) &=& \sum_s cos \chi{\cal C}_{s,-s,\mu}(p_1, k),
\\
{\Delta_T\cal C}'_{\mu}(p_1, k) &=& \sum_s sin \chi{\cal C}_{s,-s,\mu}(p_1, k)
 \label{dens1}  
\end{eqnarray} 
are correlation functions. In order to parameterize these functions, we recall the definition (\ref{rec1}) of quark-gluon-quark correlator and eq. (\ref{axi}),
concerning the gauge used. Therefore we have to take into account the available transverse four-vectors, whence it follows that
\begin{eqnarray}
{\cal C}_{\mu} &=& {\cal C}_1 p_{1\perp\mu} +\epsilon_{\mu\nu\rho\sigma}
n_-^{\nu} ({\cal C}_2 \lambda n_+^{\rho}p_{1\perp}^{\sigma} +{\cal C}_3 MS_{\bot}^{\rho}n_+^{\sigma}), \label{cmu}
\\
\Delta{\cal C}_{\mu} &=& \Delta {\cal C}p_{1\perp\mu}, \label{dcmu}
\\
\Delta_T{\cal C}_{\mu} &=& \Delta_T {\cal C} p_{1\perp\mu}, \label{dtcmu}
\\
\Delta_T{\cal C}'_{\mu} &=& \Delta_T {\cal C}' p_{1\perp\mu}. \label{dscmu}
\end{eqnarray}
Here ${\cal C}_1$, ${\cal C}_2$, ${\cal C}_3$, $\Delta {\cal C}$, $\Delta_T {\cal C}$  and $\Delta_T {\cal C}'$ are "soft" functions of $p$ and $k$. The parameterization of $\Phi^{(1)}_{\mu}$ is obtained by inserting eqs. (\ref{rec5}) and (\ref{cmu}) to (\ref{dscmu}) into eq. (\ref{fi1}). Again, as in the case of $\Gamma_0$, the Politzer theorem, of which eq. (\ref{eqd}) is a consequence, implies that renormalization effects preserve the structure of that parameterization.

\vskip 0.30in

\setcounter{equation}{0}
\renewcommand\theequation{C. \arabic{equation}}

\appendix{\large \bf Appendix C}

Here we consider the parameterization of the correlator in terms of the Dirac components, up to and including twist-3 terms. This parameterization is similar to the usual ones (Boer and Mulders, 1998; Goeke {\it et al.}, 2005), also as regards notations, except for an energy scale $\mu_0$, which we leave undetermined here, and for the twist-2, T-odd sector, which we omit because it has no place in our procedure. The scale $\mu_0$, usually set equal to the rest mass of the hadron, is determined in sects. 4 and 5, with a different result.

The parameterization reads
\begin{equation}
\Phi = 2p^+\left[(\Psi^f_E+\Psi^f_H)\delta(p^2-m^2)+ 
(\Psi^i_O+\Psi^i_H)\frac{1}{\pi(p^2-m^2+i\epsilon)}\right].\label{parm}
\end{equation}
Here 
\begin{eqnarray}
\Psi^f_E &=& \frac{\cal P}{\sqrt{2}}\{f_1 n\hspace{-0.5 em}/_+ + (\lambda 
g_{1L}+\lambda_{\perp}g_{1T})\gamma_5n\hspace{-0.5 em}/_+ + \frac{1}{2}h_{1T}\gamma_5[S\hspace{-0.65 
em}/_{\perp},n\hspace{-0.5 em}/_+]\nonumber
\\ 
&+& \frac{1}{2}(\lambda h^{\perp}_{1L}+\lambda_{\perp} h^{\perp}_{1T}) \gamma_5 [\eta\hspace{-0.5 em}/_{\perp},n\hspace{-0.5 em}/_+]\},
\\
\Psi_H &=& \frac{1}{2}\{(f^{\perp}+\lambda g^{\perp}_L\gamma_5 + \lambda_{\perp} g^{\perp}_T\gamma_5)p\hspace{-0.45 em}/_{\perp} + \frac{1}{4}\lambda_{\perp} h_T^{\perp} \gamma_5 [S\hspace{-0.65 em}/_{\perp}, p\hspace{-0.5 
em}/_{\perp}] \nonumber
\\
&+& \frac{1}{2}x M \left(e + g'_T\gamma_5 S\hspace{-0.65 em}/_{\perp} + \frac{1}{2}(\lambda h_L  + \lambda_{\perp} h_T) \gamma_5 
[n\hspace{-0.5 em}/_-,n\hspace{-0.5 em}/_+]\right)\},
\\
\Psi^i_O &=& \epsilon_{ij}S_{\bot}^i(p_{\bot}^j e_T^{\bot} + M \gamma^j f_T)+ \epsilon_{ij}{\bar S}_{\bot}^{i} p_{\bot}^j e_T^{'\bot}+\gamma_5 (xM e_L\lambda \nonumber  
\\
&+& e_T p_{\bot}\cdot S_{\bot} + e'_T p_{\bot}\cdot {\bar S}_{\bot})  + \epsilon_{ij} \gamma_i p_{\bot}^j 
(f_L^{\bot}\lambda + f_T^{\bot}\lambda_{\bot} + \gamma_5 g^{\bot}) \nonumber 
\\  
&+&\gamma_5 p\hspace{-0.45 em}/_{\bot} {\bar S}\hspace{-0.65 em}/_{\bot}h' +\frac{1}{2}\gamma_5 [\gamma_+,\gamma_-]p_{\bot}\cdot {\bar S}_{\bot}h^{'\bot}. 
\end{eqnarray}
Here $\Psi_H$ denotes the "hybrid" term, both interaction free ($\Psi_H^f$, T-even) and interaction dependent ($\Psi_H^i$, T-odd): the two terms have the same parameterization, but behave quite differently.  For the "soft" 
functions we have adopted notations similar to those employed by Goeke {\it et al.} (2005). Note, however, that in the expression of $\Psi^i_O$ the functions $f_T$, $e_T^{'\bot}$, $e'_T$, $h'$ and $h^{'\bot}$ do not appear in the parameterization proposed by those authors; on the contrary, we have not taken into account the functions $h$ and $f^{'\bot}_T$, defined by them.

\centerline{\bf References}

Abe, K., {\it et al.} (BELLE), 2006, {\it Phys. Rev. Lett.} {\bf 96}, 232002

Abe, K., {\it et al.} (E143), 1998, {\it Phys. Rev. D} {\bf 58}, 112003

Abe, K., {\it et al.} (E154), 1997a, {\it Phys. Lett. B} {\bf 405}, 180

Abe, K., {\it et al.} (E154), 1997b, {\it Phys. Rev. Lett.} {\bf 79}, 26

Adams, M.R., {\it et al.} (E665), 1993, {\it Phys. Rev. D} {\bf 48}, 5057  

Adeva, B., {\it et al.} (EMC), 1998, {\it Phys. Rev. D} {\bf 58}, 112001

Ageev, E. S., {\it et al.} (COMPASS), 2007, {\it Nucl. Phys B} {\bf 765}, 31

Airapetian, A., {\it et al.} (HERMES), 1998, {\it Phys. Lett. B} {\bf 442},  484

Airapetian, A., {\it et al.} (HERMES), 2000, {\it Phys. Rev. Lett.} {\bf 84}, 4047

Airapetian, A., {\it et al.} (HERMES), 2001, {\it Phys. Rev. D} {\bf 64}, 097101
 
Airapetian, A., {\it et al.} (HERMES), 2003, {\it Phys. Lett. B} {\bf 562}, 182

Airapetian, A., {\it et al.} (HERMES), 2005a, {\it Phys. Rev. Lett.} {\bf 94}, 012002

Airapetian, A., {\it et al.} (HERMES), 2005b, {\it Phys. Rev. D} {\bf 71},  012003

Alexakhin, V. Yu., {\it et al.} (COMPASS), 2005, {\it Phys. Rev. Lett.} 
{\bf 94}, 202002

Alekseev, M. G., {\it et al.} (COMPASS), 2010a, {\it Phys. Lett. B} {\bf 692},
240

Alekseev, M. G., {\it et al.} (COMPASS), 2010b, {\it Phys. Lett. B} {\bf 693}, 227

Anselmino, M., A. Efremov and E. Leader, 1995, {\it Phys. Rep.} {\bf 261}, 1

Anselmino, M., A. Efremov, A. Kotzinian and B. Parsamyan, 2006, {\it Phys. Rev. D} {\bf 74}, 074015

Anselmino, M., M. Boglione, U. D'Alesio, A. Kotzinian, F. Murgia, A. Prokudin and C. Turk, 2007, {\it Phys. Rev. D} {\bf 75}, 054032

Anselmino, M., M. Boglione, U. D'Alesio, S. Melis, F. Murgia and A. Prokudin, 2009a, {\it Phys. Rev. D} {\bf 79}, 054010

Anselmino, M., M. Boglione, U. D'Alesio, A. Kotzinian, S. Melis, F. Murgia, A. Prokudin and C. Turk, 2009b, {\it Eur. Phys. Jou. A} {\bf 39}, 89 

Anselmino, M., M. Boglione, U. D'Alesio, S. Melis, F. Murgia, and A. Prokudin, 2010, {\it Phys. Rev. D} {\bf 81}, 034007

Anthony, P. L., {\it et al.} (E155), 2003, {\it Phys. Lett. B} {\bf 553}, 18

Anthony, P. L., {\it et al.} (E142), 1996a, {\it Phys. Rev. D} {\bf 54}, 6620 

Anthony, P. L., {\it et al.} (E143), 1996b, {\it Phys. Rev. Lett.} {\bf 76}, 587

Arneodo, M., {\it et al.} (EMC), 1987, {\it Z. Phys. C} {\bf  34}, 277  

Artru, X., and M. Mekhfi, 1990, {\it Z. Phys. C} {\bf 45}, 669

Ashman, J., {\it et al.} (EMC), 1988, {\it Phys. Lett. B} {\bf 206}, 364

Ashman, J., {\it et al.} (EMC), 1989, {\it Nucl. Phys B} {\bf 328}, 1

Ashman, J., {\it et al.} (EMC), 1991, {\it Z. Phys. C} {\bf  52}, 361  

Avakian, H., P. Bosted, V. Burkert and L. Elouadrhiri, 2005, {\it AIP Conf. Proc.} {\bf 792}, 945

Avakian, H., A. V. Efremov, K. Goeke, A. Metz, P. Schweitzer and T. Tecknentrup, 2008a, {\it Phys. Rev. D} {\bf 77}, 014023

Avakian, H., A. V. Efremov, P. Schweitzer and F. Yuan, 2008b, {\it Phys. Rev. D} {\bf 78}, 114024


Bacchetta, A., D. Boer, M. Diehl and P. J. Mulders, 2008, {\it JHEP} {\bf 0808:023} 

Belitsky, A. V., X. Ji and F. Yuan, 2003, {\it Nucl. Phys. B} {\bf 656}, 165

Berger, E. L., and S. J. Brodsky, 1979, {\it Phys. Rev. Lett.} {\bf 42}, 940

Bilal, A., {\it et al.}, 1991, {\it Nucl. Phys. B} {\bf 355}, 549

Bluemlein, J., and A. Tkabladze, 1999, {\it Nucl. Phys. B} {\bf 553}, 427

Boer, D. and P. J. Mulders, 1998, {\it Phys. Rev. D} {\bf 57}, 5780

Boer, D., P. J. Mulders and O. V. Teryaev, 1998, {\it Phys. Rev. D} {\bf 57}, 3057

Boer, D., 1999, {\it Phys. Rev. D} {\bf 60}, 014012

Boer, D., R. Jakob and P. J. Mulders, 2000, {\it Nucl. Phys. B} {\bf 564}, 471

Boer, D., S. J. Brodsky and D.-S. Huang, 2003a, {\it Phys. Rev. D} {\bf 67}, 054003

Boer, D., P. J. Mulders and F. Pijlman, 2003b, {\it Nucl. Phys. B} {\bf 667}, 201

Boer, D., 2009, {\it Nucl. Phys. B} {\bf 806}, 23

Boffi, S., A. V. Efremov, B. Pasquini and P. Schweitzer, 2009, {\it Phys. Rev. D} {\bf 79}, 094012  

Bomhof, C.J., P.J. Mulders and F. Pijlman, 2004, {\it Phys. Lett. B} {\bf 596}, 277

Bravar, A. {\it et al.} (SMC), 1999, {\it Nucl. Phys. B (Proc. Suppl.)} {\bf 79}, 520

Bressan, A. (COMPASS), 2007, {\it Munich Deep inelastic scattering}, Vol. $1^{st}$, pag. 583

Brodsky, S. J., D. S. Hwang and I. Schmidt, 2002a, {\it Phys. Lett. B} {\bf 530}, 99

Brodsky, S. J., D. S. Hwang and I. Schmidt, 2002b, {\it Nucl. Phys. B} {\bf 642}, 344

Brodsky, S. J., D. S. Hwang and I. Schmidt, 2003, {\it Int. Jou. Mod. Phys. A} {\bf 18}, 1327

Bunce, G., N. Saito, J. Soffer and W. Vogelsang, 2000, {\it Ann. Rev. Nucl. Part. Sci.} {\bf 50} 525

Burkardt, M., and B. Hannafious, 2008, {\it Phys. Lett. B} {\bf 658}, 130

Burkhardt, H., and W. N. Cottigham, 1970, {\it Ann. Phys.} {\bf 56}, 453


Cahn, R.N., 1978, {\it Phys. Lett. B} {\bf 78}, 269

Cahn, R.N., 1989, {\it Phys. Rev. D} {\bf 40}, 3107

Catani, S., B. R. Webber and G. Marchesini, 1991a, {\it Nucl. Phys. B} {\bf 349}, 635

Catani, S., B. R. Webber, F. Fiorani and G. Marchesini, 1991b, {\it Nucl. Phys. B (Proc. Suppl.)} {\bf 23}, 123

Collins, J. C., 1993, {\it Nucl. Phys. B} {\bf 396}, 161

Collins, J. C., A. V. Efremov, K. Goeke, M. Grosse Perdekamp, S. Menzel, B. Meredith, A. Metz and P. Schweitzer, 2006, {\it Phys. Rev. D} {\bf 73}, 094023 

Collins, J. C., 1998, {\it Phys. Rev. D} {\bf 57}, 3051

Collins, J. C., 1989, {\it Perturbative QCD}, A.H. Mueller ed., World Scientific, Singapore 

Collins, J. C.,  D. E. Soper and G. Sterman, 1988, {\it Adv. Ser. Direct. High Energy Phys.} {\bf 5}, 1 

Collins, J. C., and D. E. Soper, 1981, {\it Nucl. Phys. B} {\bf 193}, 381

Collins, J. C., and D. E. Soper, 1982, {\it Nucl. Phys. B} {\bf 197}, 446

Collins, J. C., 2002, {\it Phys. Lett. B} {\bf 536}, 43

Collins, J. C., and J. Qiu, 2007, {\it Phys. Rev. D} {\bf 75}, 114014

Conway, J.S., {\it et al.}, 1989, {\it Phys. Rev. D} {\bf 39}, 92


DeRujula, A., et al., 1971, {\it Nucl. Phys. B} {\bf 35}, 365

Di Salvo, E., 2007a, {\it Int. Jou. Mod. Phys. A} {\bf 22}, 2145

Di Salvo, E., 2007b, {\it Mod. Phys. Lett. A} {\bf 22}, 1787

Di Salvo, E., 2003, {\it Int. Jou. Mod. Phys. A} {\bf 18}, 5277

Di Salvo, E., 2002, {\it Nucl. Phys. A} {\bf 711}, 76

Di Salvo, E., 2001, {\it Eur. Phys. Jou. C} {\bf 19}, 503

Diefenthaler, M. (HERMES), 2005, {\it AIP Conf. Proc.} {\bf 792}, 933

Dokshitzer, Yu. L., D. I. Dyakonov and S. I. Troyan, 1980, {\it Phys. Rep.} {\bf 58}, 269


Ellis, R. K., W. Furmanski and R. Petronzio, 1982, {\it Nucl. Phys. B} 
{\bf 207}, 1 

Ellis, R. K., W. Furmanski and R. Petronzio, 1983, {\it Nucl. Phys. B} 
{\bf 212}, 29 

Efremov, A., and A. Radyushkin, 1981, {\it Theoretical and Mathematical Physics} {\bf 44}, 774

Efremov, A., and O. Teryaev, 1984, {\it Yad. Fiz.} {\bf 39}, 1517

Efremov, A. V., K. Goeke and P. Schweitzer, 2006a, {\it Phys. Rev. D} {\bf 73}, 094025

Efremov, A. V., K. Goeke and P. Schweitzer, 2006b, {\it Czech. J. Phys.} 
{\bf 56}, F181

Efremov, A. V., and O. Teryaev, 1985, {\it Phys. Lett. B} {\bf 150}, 383

Efremov, A. V., P. Schweitzer, O. V. Teryaev and P. Zavada, 2009, {\it Phys. Rev. D} {\bf 80}, 014021

Efremov, A. V., O. V. Teryaev and E. Leader, 1997 {\it Phys. Rev. D} {\bf 55},
4307


Falciano, S., {\it et al.} (NA10), 1986, {\it Z. Phys. C - Particles and Fields} {\bf 31}, 513


Goeke, K., A. Metz and M. Schlegel, 2005, {\it Phys. Lett. B} {\bf 618}, 90

Guanziroli, M., {\it et al.} (NA10), 1988, {\it Z. Phys. C - Particles and Fields} {\bf 37}, 545
  

Hawranek, P. (PANDA), 2007, {\it Int. Jou. Mod. Phys. A} {\bf 22}, 574


Jaffe, R. L., and X. Ji, 1991a, {\it Phys. Rev. Lett.} {\bf 67}, 552 

Jaffe, R. L., and X. Ji, 1991b, {\it Phys. Rev. D} {\bf 43}, 724

Jaffe, R. L., and X. Ji, 1992, {\it Nucl. Phys. B} {\bf 375}, 527

Ji, X., and F. Yuan, 2002, {\it Phys. Lett. B} {\bf 543}, 66

Ji, X., J. W. Qiu, W. Vogelsang and F. Yuan, 2006a, {\it Phys. Rev. Lett.} 
{\bf 97}, 082002

Ji, X., J. W. Qiu, W. Vogelsang and F. Yuan, 2006b, {\it Phys. Rev. D} {\bf 73}, 094017

Ji, X., J. W. Qiu, W. Vogelsang and F. Yuan, 2006c,  {\it Phys. Lett. B} {\bf 638}, 178


Kodaira, J., and H. Yokoya, 2003, {\it Phys. Rev. D} {\bf 67}, 074008

Kogut, J. B., and D. E. Soper, 1970, {\it Phys. Rev. D} {\bf 1}, 2901

Koike, Y., W. Vogelsang and F. Yuan, 2008, {\it Phys. Lett. B} {\bf 659}, 878

Kotzinian, A. M., 1995, {\it Nucl. Phys. B} {\bf 441}, 234

Kotzinian, A. M., and P. J. Mulders,  1996, {\it Phys. Rev. D} {\bf 54}, 1229


Lenisa, P. (PAX), 2005, {\it AIP Conf. Proc.} {\bf 792}, 1023

Levelt, J., and P. J. Mulders, 1994, {\it Phys. Rev. D} {\bf 49}, 96

Leader, E., 2004, {\it Phys. Rev. D} {\bf 70}, 054019


McGaughey, P. L., {\it et al.} (E772), 1994, {\it Phys. Rev. D} {\bf 50}, 3038

Mulders, P. J., and R. D. Tangerman, 1996, {\it Nucl. Phys. B} {\bf 461}, 197


Peign\'e, S., 2002, {\it Phys. Rev. D} {\bf 66}, 114011

Politzer, H. D., 1980, {\it Nucl. Phys. B} {\bf 172}, 349


Qiu, J., and G. Sterman, 1991, {\it Phys. Rev. Lett.} {\bf 67}, 2264

Qiu, J., and G. Sterman, 1992, {\it Nucl. Phys. B} {\bf 378}, 52 
 
Qiu, J., and G. Sterman, 1998, {\it Phys. Rev. D} {\bf 59}, 014004

Qiu, J., 1990, {\it Phys. Rev. D} {\bf 42}, 30


Ralston, J., and D. E. Soper, 1979, {\it Nucl. Phys B} {\bf 152}, 109

Rogers, T. C., 2007, {\it 8th International Symposium on Radiative Corrections (RADCOR)}, October 1-5, arXiv:hep-ph/0712.1195. See also references therein.


Sivers, D. W., 1990, {\it Phys. Rev. D} {\bf 41}, 83

Sivers, D. W., 1991, {\it Phys. Rev. D} {\bf 43}, 261

Sivers, D., 2006, {\it Phys. Rev. D} {\bf 74}, 094008
 
Soffer, J., and P. Taxil, 1980, {\it Nucl. Phys. B} {\bf 172}, 106

Soffer, J., 1995, {\it Phys. Rev. Lett.} {74}, 1292

Sterman, G., 2005, {\it Contribution to Elsevier Encyclopedia of Mathematical Physics}, YITP-SB-05-27, hep-ph/0512344 


Tangerman, R. D., and P. J. Mulders, 1995, {\it Phys. Rev. D} {\bf 51}, 3357

Towell, R. S., {\it et al.} (E866), 2001, {\it Phys. Rev. D} {\bf 64}, 052002

Yun, J., {\it et al.} (CLAS), 2003, {\it Phys. Rev. C} {\bf 67}, 055204

Zheng, X., {\it et al.} (E-99-117), 2004, {\it Phys. Rev. C} {\bf 70}, 065207

Zhu, L. Y., {\it et al.} (E866), 2007, {\it Phys. Rev. Lett.} {\bf 99}, 082301




\end{document}